%% file: Main_CVPR.tex
\documentclass[10pt,twocolumn,letterpaper]{article}

\usepackage[pagenumbers]{cvpr}      

\usepackage{graphicx}
\usepackage{amsmath}
\usepackage{amssymb}
\usepackage{booktabs}

\usepackage[pagebackref,breaklinks,colorlinks]{hyperref}

\usepackage{url}

\usepackage[utf8]{inputenc} 
\usepackage[T1]{fontenc}    
\usepackage{booktabs}       
\usepackage{amsfonts}       
\usepackage{nicefrac}       
\usepackage{microtype}      
\usepackage{xcolor}         
\usepackage{float}
\usepackage{amsmath}
\usepackage{multirow}
\usepackage{xspace}
\usepackage[ruled,vlined,linesnumbered,noresetcount]{algorithm2e}
\usepackage{stfloats}
\usepackage{subfiles}
\usepackage{wrapfig}
\usepackage{amssymb}
\usepackage{pifont}
\newcommand{\cmark}{\ding{51}}%
\newcommand{\xmark}{\ding{55}}%
\newif\ifshowcomments
 \showcommentstrue
\ifshowcomments
\newcommand{\mynote}[2]{\fbox{\bfseries\sffamily\scriptsize{#1}}
{\small$\blacktriangleright$\textsf{\emph{#2}}$\blacktriangleleft$}}
\else
\newcommand{\mynote}[2]{}
\fi

\SetKwFunction{FTrain}{Train}

\usepackage[capitalize]{cleveref}
\crefname{section}{Sec.}{Secs.}
\Crefname{section}{Section}{Sections}
\Crefname{table}{Table}{Tables}
\crefname{table}{Tab.}{Tabs.}

\def\cvprPaperID{5803} 
\def\confName{CVPR}
\def\confYear{2022}

\begin{document}
\newcommand{\alg}{\textsc{MEGA}\xspace}

\title{MEGA: Model Stealing via Collaborative Generator-Substitute Networks}

\author{Chi Hong\\
Delft University of Technology\\
{\tt\small C.Hong@tudelft.nl}
\and
Jiyue Huang\\
Delft University of Technology\\
{\tt\small j.huang-4@tudelft.nl}
\and
Lydia Y. Chen\\
Delft University of Technology\\
{\tt\small lydiaychen@ieee.org}
}
\maketitle

\begin{abstract}

Deep machine learning models are increasingly deployed in the wild for providing services to users. Adversaries may steal the knowledge of these valuable models by training substitute models according to the inference results of the targeted deployed models. Recent data-free model stealing methods are shown effective to extract the knowledge of the target model without using real query examples, but they assume rich inference information, e.g., class probabilities and logits. However, they are all based on competing generator-substitute networks and hence encounter training instability. In this paper we propose a data-free model stealing framework, \alg, which is based on collaborative generator-substitute networks and only requires the target model to provide label prediction for synthetic query examples. The core of our method is a model stealing optimization consisting of two collaborative models (i) the substitute model which imitates the target model through the synthetic query examples and their inferred labels and (ii) the generator which synthesizes images such that the confidence of the substitute model over each query example is maximized. We propose a novel coordinate descent training procedure and analyze its convergence. We also empirically evaluate the trained substitute model on three datasets and its application on black-box adversarial attacks. Our results show that the accuracy of our trained substitute model and the adversarial attack success rate over it can be up to 33\% and 40\% higher than state-of-the-art data-free black-box attacks.

\end{abstract}

\section{Introduction}
\label{sec:intro}
\input{sections/introduction}

\section{Related work}
\label{sec:rela}
\input{sections/related_work}

\section{Methodology}
\label{sec:method}
\input{sections/methodology}

\section{Evaluation}
\label{sec:exp}
\input{sections/experiments}

\section{Conclusion}
It's challenging to design adversarial attacks without knowing the target model parameters nor access to real-world data. In this paper, we propose a novel and effective data-free model stealing framework, \alg, a collaborative generator-substitute networks, which aims to steal the knowledge of the target model using synthetically generated queries.  
To collaboratively steal the target model, on the one hand, the generator synthesizes data such that the inference confidence of the substitute model on synthetic data is maximized. On the other hand, the substitute model minimizes the cross entropy loss between its predictions and the target model on synthetic data. 
To ensure that the algorithm converges steadily and efficiently, we train the generator and the substitute model in a coordinate descent fashion by repetitively exploiting every set of synthetic images and theoretically prove its convergence. We empirically demonstrate that \alg effectively steals the target model using a small number of queries for three datasets. Under various adversarial scenarios, we show that the substitute model achieves 19\%--33\% higher accuracy and up to 40\% higher adversarial attack success rate than state-of-the-art data-free model stealing black-box attacks.

{\small
\bibliographystyle{ieee_fullname}
\bibliography{egbib}
}

\clearpage

\appendix

\input{sections/appendix}

\end{document}

%% file: sections/introduction.tex
Machine learning has been applied in multiple fields like computer vision and natural language processing to establish various services. Emerging intelligent services, such as Google translate and optical character recognition~\cite{googlecloud}, are increasingly powered by trained deep models. 
Users can access these services by sending queries through APIs to get outputs, for instance, the class labels of the queried images. 
While such open access to deployed models greatly eases users' experience, it may also open up various vulnerability issues related to model stealing \cite{truong2021data, zhou2020dast, Kariyappa21MAZE}. Adversaries may make use of such an access to steal the knowledge of the model and create a copy of it, named a substitute model. Malicious parties can then monetize such substitute models through hosting inference services~\cite{DBLP:conf/iclr/KrishnaTPPI20}.
In addition, they can be applied to further launch  attacks such as crafting adversarial examples to fool the deployed models~\cite{Ebrahimi18white, fang2019data,truong2021data} and conducting membership attacks~\cite{yeom2018privacy}.

In model stealing attacks, an adversary tries to train a substitute model that mimics the inference results of the target model~\footnote{We interchangeably use the terms of target and deployed models.}. Using APIs, the adversary can get the corresponding inference result of any given query input, e.g., image and their classification label. 
Such query-result pairs are used as an input to train the substitute model which is a imitation of the target model. To launch a model stealing attack, the adversary needs examples from a surrogate dataset which is semantically similar to the original training dataset to query the target model~\cite{orekondy2019knockoff}. However, in most real-world scenarios, curating such a surrogate dataset can be difficult because the data from privacy- and business-sensitive domains, e.g., banks and medical institutions, may not be accessible by attackers. Furthermore, it is time-consuming and expensive to collect sufficient real data examples to query the target model so as to train the substitute model.

Recognizing the limited availability of surrogate data for adversarial queries, recent studies~\cite{truong2021data, Kariyappa21MAZE} propose data-free model stealing methods. In these methods, a generator is designed to synthesize query examples as the input to the target model. 
To train their generators, the corresponding inference probabilities~\cite{truong2021data, Kariyappa21MAZE} or logits~\cite{truong2021data} of the input examples from the target model are required to do gradient approximation~\cite{polyak1987introduction}, because adversaries have no access to the actual gradient of the target model. However, the need of inference probabilities and logits limits their applicability to real-world problems where often only the label inference of query data is provided by the target model. Moreover, aforementioned approaches train the generator and substitute model in competition, similar to the principle of Generative Adversarial Networks (GANs). Specifically, the substitute model imitates the prediction of the target but the generator maximizes the disagreement between the predictions of the target and the substitute model. The inherent training instability from GANs not only persists but also aggravates in existing data-free model stealing methods. 



In this paper, we aim to steal a model in a challenging adversarial scenario in which there is a black-box target model that only provides label prediction without any additional information on class probabilities or logits. At the same time, adversaries don't have any real data for querying the target model. We propose a novel data-free model stealing framework, \alg, which only needs the label inferences of synthetically generated query data to learn the substitute model. We design the generator of \alg from a novel prospective which is different from min-max GANs to avoid unstable training. In \alg, We design an architecture with collaborative generator-substitute networks to steal the target model using synthetically generated images and predicted labels from the target model. To collaboratively steal the target model, the substitute model minimizes the cross entropy loss between its and the target model's predictions while the generator indirectly minimizes the loss of substitute model by maximizing the inference confidence of the substitute model on synthetic data. For ensuring that the algorithm converges steadily and efficiently, we propose to train the generator and the substitute model in a coordinate descent way and exploit the synthetic datasets using a given set of noise seeds in multiple training rounds. 

The contributions of this paper are shown in the following:

\begin{itemize}
    \item In section~\ref{sec:method}, we propose a first kind of collaborative generator-substitute networks for data-free model stealing framework, \alg, which is based on a novel collaborative training principle and a coordinate descent training procedure. We provide the theoretical analysis on the convergence of \alg. 
    \item In section~\ref{sec:exp}, we extensively compare \alg with state-of-the-art data-free model stealing approaches and their applications on black-box adversarial attacks. Our empirical results show that \alg can effectively steal models and further fool the target model using significantly fewer synthetic queries. 
\end{itemize}

%% file: sections/related_work.tex

In this section, we first provide an overview on model stealing methods with detailed comparison of methods that are closely related to data free stealing. We then discuss the  adversarial attack that can leverage the stolen models. 

\textbf{Model stealing.}
The goal of model stealing is to distill the knowledge from a deployed model (target model). Specifically, it is to train a highly similar substitute model \cite{krishna2019thieves,jagielski2020high, chandrasekaran2020exploring,zhou2020dast,truong2021data,juuti2019prada,orekondy2019knockoff}.
A successful substitute model is able to obtain the implicit mapping function (or knowledge, in high level) of the target model by different network structures~\cite{Kariyappa21MAZE,Orekondy19knockoff}. 
There are two types of model stealing methods depending on whether the attackers are able to access the real training data (or part of it). In the case when real data is available, knowledge distilling ~\cite{hinton2015distilling} extracts the knowledge of the target model.
The key idea is that the substitute model is trained by class probabilities of the target using part of the dataset. The class probabilities (so called ``soft targets'') are produced by inferring the target model. By this means, the knowledge of the complex target model can be transferred to a lightweight substitute model in order to reduce inference cost. When real data is not available (black-box) for inference, attackers can only imitate the target model through querying synthetic examples~\cite{MicaelliS19,krishna2019thieves,zhou2020dast,Kariyappa21MAZE}. For example,  \cite{krishna2019thieves} uses randomly generated sentences for querying NLP models, but lacks generalization beyond NLP tasks. 
\begin{table}[h]
\renewcommand\arraystretch{1.25}
\centering
\caption{Existing data-free model stealing methods.}
\label{tab:exmethod}
\resizebox{1\columnwidth}{!}{%
\begin{tabular}{c|ccccc} 
\toprule
\textbf{Method} & Competitive training & $\mathcal{L}_{\mathcal{S}}$ & $\mathcal{L}_{\mathcal{G}}$ & label-only & probability-only \\ 
\midrule
DFME~\cite{truong2021data}     &          \cmark       &    $\left\|\mathcal{T}(x)-\mathcal{S}(x)\right\|_{1}$       &           $- \left\|\mathcal{T}(x)-\mathcal{S}(x)\right\|_{1}$ & \xmark    &     	\cmark
       \\ 
\bottomrule
DaST~\cite{zhou2020dast}  &          \cmark       &           $\mathrm{CE}(\mathcal{T}(x), \mathcal{S}(x))$           & $e^{-\mathrm{CE}(\mathcal{T}(x), \mathcal{S}(x))}$ &    \cmark      &   \cmark  \\ 
\bottomrule

\end{tabular}
}

\end{table}

In the exiting black-box model stealing methods~\cite{Kariyappa21MAZE,truong2021data,zhou2020dast}, the core is a competing generator-substitute networks. The model parameters and architecture of the target model are unknown to adversaries. A generator produces synthetic examples to query the target model, whereas the substitute model tries to imitate/steal the target model through the synthetic queries. 
Specifically, MAZE~\cite{Kariyappa21MAZE} and DFME~\cite{truong2021data} rely on a gradient approximation method~\cite{polyak1987introduction} to estimate the gradient of the target model. The estimated gradient is applied to train their generator. DaST~\cite{zhou2020dast} regards the output of the target model as a constant vector and doesn't need gradient approximation to train the generator. 

Let‘s further zoom into the details of DaST and DFME, summarized in Table~\ref{tab:exmethod}. 
The substitute model tries to mimic the prediction of the target model while the generator maximizes the disagreement between the predictions of the target and the substitute model. Two models are trained competitively. Specifically, DFME relies on L1 norm loss between the prediction of image $x$, by the target model $\mathcal{T}(x)$, and substitute model $\mathcal{S}(x)$, whereas DaST uses the cross entropy loss between the target and substitute models, i.e., $\mathrm{CE}(\mathcal{T}(x),\mathcal{S}(x))$. Since the gradient approximation requires the target model to provide the inference probabilities or logits for query examples, DFME can not be applied on scenario where only inference labels are provided by the target model. 
Such a competitive training scheme makes the loss and the accuracy of the substitute model oscillate a lot during the training~\cite{farnia2020gans}. Especially, when there is no real data to train the generator in data-free model stealing, the random seeds input to the generator further aggravates the oscillation.

\textbf{Adversarial attacks. } 
Adversarial attacks aim to generate visual indistinguishable adversarial examples to fool target models. Attacks can be under white-box setting~\cite{dong2018boosting, Goodfellow:2015:FGSM, Kurakin:17:BIM, papernot2016limitations,carlini2017towards,szegedy2013intriguing}, or black-box attack settings~\cite{ilyas2018black}. As the crucial difference among them is that white-box attackers can access the target model, most white-box attack methods can be applied for black-box setting using a substitute model. Thus, in the following, we assume that substitute models are available in black-box setting so that no separate discussion on white-box or black-box attacks is needed. 
The gradient-based approaches~\cite{dong2018boosting, Goodfellow:2015:FGSM, Kurakin:17:BIM, papernot2016limitations} are extensively explored. FGSM~\cite{Goodfellow:2015:FGSM} is a one-step attack that generates the noise according to the gradient signs of the examples and adds the noise back to the normal examples to obtain the corresponding adversarial examples. BIM~\cite{Kurakin:17:BIM} is an enhanced iterative version of FGSM while MI-FGSM ~\cite{dong2018boosting} elaborates the momentum on BIM for higher transferability. 
PGD~\cite{Madry:18:PGD, croce2020reliable}, the state-of-the-art iterative version of FGSM, initializes the example at a random point and randomly restarts in each iteration. The substitute model obtained from \alg is compatible and applicable with these gradient-based adversarial methods.



%% file: sections/methodology.tex
\label{sec:method}

In this section, we first define the adversarial model and then introduce our data-free model stealing framework, \alg. We focus on black-box setting where we have neither knowledge on the model parameters of the target model nor possession of any real data to query it. 

\subsection{Adversarial Models}
\label{subsec:codef}
In this paper, a target model $\mathcal{T}$ that conducts classification tasks is deployed. The target model $\mathcal{T}$ has a black-box setting that the parameters and the architecture are unknown to adversaries. The goal of adversaries is to steal the knowledge of $\mathcal{T}$ by training a substitute model $\mathcal{S}$ which can be regarded as a cloned model of $\mathcal{T}$. We assume that adversaries do not have any surrogate dataset which is semantically similar to the original training dataset of $\mathcal{T}$ and they do model stealing without real data. Although attackers can't access the target model, they can query it to get the inference results. We consider a hard adversarial scenario that $\mathcal{T}$ only provides a label prediction for each query without any additional information on its class probabilities or logits. We call this scenario as \textit{label-only} scenario. 

\begin{figure}[h]
  \centering
  \includegraphics[width=0.48\textwidth]{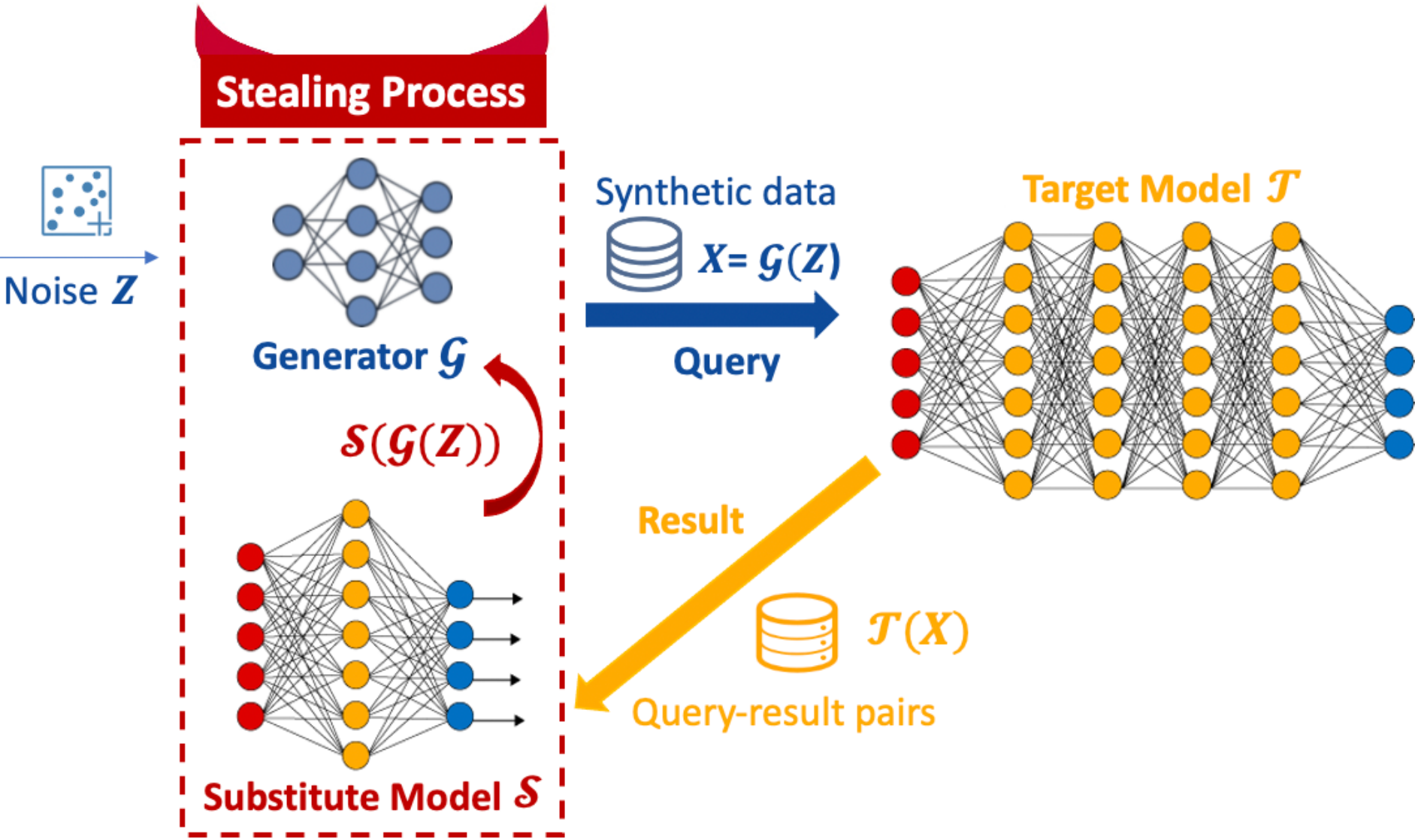}
  \caption{\alg framework: data-free model stealing process.}
  \label{ms_based_df_attack}
\end{figure}

\subsection{\alg}
In the following, we propose our algorithm \alg which can steadily steal the knowledge from $\mathcal{T}$ and train a accurate substitute model $\mathcal{S}$ even in the label-only scenario. Here, we assume $\mathcal{T}$ to be a model of classification task, with $N$ classes.

\textbf{Notation:}
In this data-free approach, a generator $\mathcal{G}$ is required to produce synthetic examples to query $\mathcal{T}$. A set of noise vectors $Z = \{z_1, ..., z_M\}$ where $M$ is the number of noise vectors, is generated as the input seeds of $\mathcal{G}$. The corresponding synthetic examples $X = \{x_i \mid x_i = \mathcal{G}(z_i), i = 1, ..., M\}$ are generated by the generator. $X$ is used to query the target model so that we can obtain the query results $\mathcal{T}(X) = \{\mathcal{T}(x_i) \mid i = 1, ..., M\}$. We use $\mathcal{P}(x)$ and $\mathcal{S}(x)$ to respectively denote the output probability vector of the target model $\mathcal{T}$ and the substitute model $\mathcal{S}$ over an input example $x$. 
Specifically, $\mathcal{T}(x)$ is the one-hot vector of the predicted label of an input $x$, where $\mathcal{T}_j(x)$ denotes the $j$-th ($j = 1, ..., N$) element of the output. The model parameters of $\mathcal{S}$ and $\mathcal{G}$ are respectively represented as $\theta_s$ and $\theta_g$.

\textbf{Architecture}
The architecture of proposed model stealing framework, \alg, is shown in Figure~\ref{ms_based_df_attack}. 
In the stealing process, the generator $\mathcal{G}$ first generates synthetic examples $X=\mathcal{G}(Z)$ from noise vectors $Z$ to query $\mathcal{T}$ and get the output  $\mathcal{T}(X)$. Next, $X$ associated with $\mathcal{T}(X)$ is fed into $\mathcal{S}$ for training. The output of the substitute model $\mathcal{S}(\mathcal{G}(Z))$ is then used to update $\mathcal{G}$ so as to generate better synthetic examples. In the framework, $\mathcal{S}$ and $\mathcal{G}$ are two networks that work collaboratively rather than competitively. Their optimization objectives are shown in the following.

\textbf{The optimization objective of the substitute model:} Since $\mathcal{S}$ is a substitute of $\mathcal{T}$, their outputs are expected to be as similar as possible. 
Inspired by knowledge distillation ~\cite{hinton2015distilling}, $\mathcal{S}$ imitates the outputs of $\mathcal{T}$ through cross entropy loss. The loss function of $\mathcal{S}$ over a query example $x$ can be defined as the following equation:
\begin{equation}
\label{loss_s}
    \mathcal{L}_{\mathcal{S}}  = \mathrm{CE}(\mathcal{T}(x), \mathcal{S}(x)) = - \sum_{i=1}^{N} \mathcal{T}_{i}(x) \log \mathcal{S}_{i}(x),
\end{equation}
where $\mathrm{CE}$ denotes cross entropy. 
After training $\mathcal{S}$ using $X$ and $\mathcal{T}(X)$, we can steal the knowledge of $\mathcal{T}$ because $\mathcal{S}$ learns the mapping of $\mathcal{T}$.

\textbf{The optimization objective of the generator:} The generator $\mathcal{G}$ is responsible for generating synthetic examples to query the target model. Here, we explain how $\mathcal{G}$ steers its synthesizing direction. If we use a real-world example $\bar{x}$ to query $\mathcal{T}$ and assume that $\mathcal{T}$ is well trained, the confidence of model $\mathcal{T}$'s output probability vector $\mathcal{P}(\bar{x})$, i.e., the biggest element of  $\mathcal{P}(\bar{x})$, is expected to be high. In our attack setting, we need to synthesize query that resembles the query of real world data. Given a synthetic example $x = \mathcal{G}(z)$, where $z$ is a noise vector, if the confidence of $\mathcal{T}(x)$ is high, $x$ is regarded as a high-quality query example~\cite{li2006confidence,zhu2010confidence}. Consequently, the objective of $\mathcal{G}$ is to generate a synthetic example $x$ that maximizes its confidence over model $\mathcal{T}$. However, we can't straightforwardly make use of maximizing the confidence over $\mathcal{T}$ to be the optimization goal for training $\mathcal{G}$. The reason is that the backpropagation of this optimization goal requires the gradient information of $\mathcal{T}$, but the model parameters of the target model are not accessible in our setting. Besides, in label-only scenario, we can't estimate the gradient of $\mathcal{T}$ using gradient approximation methods, e.g.,  the method of forward differences~\cite{polyak1987introduction,truong2021data, Kariyappa21MAZE}. Since $\mathcal{S}$ mimics the outputs of $\mathcal{T}$ in the training process, the outputs of $\mathcal{S}$  gradually draw closer to the outputs of $\mathcal{T}$ 
given the same inputs \cite{zhou2020dast}. Thus, we use $\mathcal{S}(x)$ to approximate $\mathcal{P}(x)$ and define the loss function of $\mathcal{G}$ as:
\begin{equation}
\label{loss_g}
    \mathcal{L}_{\mathcal{G}} = - \left\{\log \mathcal{S}_k \left(\mathcal{G}(z)\right) \mid \forall_{j}: \mathcal{S}_j\left(\mathcal{G}(z)\right) \leq \mathcal{S}_k\left(\mathcal{G}(z)\right)\right\},
\end{equation}
where $\mathcal{S}_j(\mathcal{G}(z))$ represents the $j$-th element of the output probability vector $\mathcal{S}(\mathcal{G}(z))$. With this loss, for an input example $x=\mathcal{G}(z)$, we maximize the value of the $k$-th element of $\mathcal{S}(x)$ where $k$ is the index of the biggest element. According to Eq.~(\ref{loss_g}), we can see that in our approach, updating $\mathcal{G}$ only requires the gradient of the substitute model. It's applicable in the label-only scenario. Actually, training $\mathcal{G}$ using $\mathcal{L}_{\mathcal{G}}$ also indirectly decreases the loss of $\mathcal{S}$ shown in Eq.~(\ref{loss_s}). We demonstrate this in Appendix~\ref{sec:lemma1}. $\mathcal{S}$ and $\mathcal{G}$ are trained collaboratively to steal the target model rather than a min-max game which is applied in existing methods~\cite{zhou2020dast, truong2021data, Kariyappa21MAZE}. 


\textbf{Stealing algorithm:}
In Algorithm~\ref{alg:code}, given a set of noise seeds $Z$, we show how \alg exploits synthetic data examples and trains the substitute model. In the training function, we have multiple rounds to iteratively train $\mathcal{S}$ and $\mathcal{G}$. Specifically, the noise vectors are  fed into $\mathcal{G}$ for generating a set of synthetic examples $X$ at the beginning of each round. We then use the synthetic examples to query $\mathcal{T}$ and obtain the query-result pairs $(X, \mathcal{T}(X))$. We apply mini-batch training in the algorithm. For the loss $\mathcal{L}_{\mathcal{G}}$, we use $\mathcal{S}(x)$ to be the approximation of $\mathcal{P}(x)$. Thus, in each round, $\mathcal{S}$ is firstly updated to imitate $\mathcal{T}$. To ensure that after training the substitute model (line 7-11), $\mathcal{S}$ is as close as to $\mathcal{T}$. Each query-result pair is not just used to update one first order stochastic optimization step (e.g., one SGD step~\cite{goodfellow2016deep}) of $\mathcal{S}$. Instead, the algorithm traverses $(X, \mathcal{T}(X))$ multiple times to train $\mathcal{S}$ within a round until the loss doesn't significantly decrease anymore. The local optima of $\mathcal{S}$ for $(X, \mathcal{T}(X))$ thus can be found after training in this way (empirically validated in Appendix~\ref{appendix:justify}). In each round, the algorithm also traverses $Z$ to train $\mathcal{G}$ (line 12-15). The updated $\mathcal{G}$ is then applied to produce new synthetic examples in the next round and the new synthetic examples thus have higher inference confidence (see Appendix~\ref{appendix:justify} for empirical evidence). We explicitly exploit the input noise set $Z$ for multiple rounds till the convergence to the local optima. 
It is important to note that \alg proceeds in a coordinate descent way because training $\mathcal{S}$ (line 7-11) minimizes $\mathcal{L}_{\mathcal{S}}$ while training $\mathcal{G}$ (line 12-15) also indirectly decreases $\mathcal{L}_{\mathcal{S}}$ (see Appendix~\ref{sec:lemma1} for proof). The algorithm continuously reduces the loss of substitute model in each round. This coordinate descent training way ensures that \alg converges steadily and efficiently.

\begin{algorithm}[h]  
  \caption{\alg} 
  \label{alg:code}
    Generate a set of noise vectors $Z = \{z\}$ \\
    $\mathcal{S} =$ Train($Z$) \\
    \SetKwProg{Fn}{Function}{:}{}
    \Fn{\FTrain{$Z$}}
    {
        \For{number of rounds}
        {
            Generate examples $X = \mathcal{G}(Z)$ using the generator\\
            Get the query results $\mathcal{T}(X)$ from the target model \\
            \tcp{train the substitute model}
            \For{number of iterations}
            {
                Sample a batch $\tilde{X}$ from $X$\\
                Get the corresponding $\mathcal{T}(\tilde{X})$\\
                Get the output of the substitute model $\mathcal{S}(\tilde{X})$\\
                Update $\theta_s$ by minimizing $\mathcal{L}_{\mathcal{S}}$ using first order stochastic optimization\\
            }
            \tcp{train the generator}
            \For{number of iterations}
            {
                Sample a batch $\tilde{Z}$ from $Z$\\
                Get the output of the substitute model $\mathcal{S}(\mathcal{G}(\tilde{Z}))$\\
                Update $\theta_g$ by minimizing $\mathcal{L}_{\mathcal{G}}$ using first order stochastic optimization\\
            }
        }
        \KwRet the trained substitute model $\mathcal{S}$ \\
    }
\end{algorithm}

\subsection{Analysis}
\label{subsec:analysis}

\textbf{Convergence:} 
\label{subsec:conv}In Algorithm~\ref{alg:code}, $\mathcal{S}$ and $\mathcal{G}$ are updated iteratively in each round $t$. Let $\theta_s^{(t)}$ and $\theta_g^{(t)}$ be the model parameters of $\mathcal{S}$ and $\mathcal{G}$ after the training of round $t$. To analyze the convergence of Algorithm~\ref{alg:code}, we make the following assumptions according to the stealing algorithm of \alg and the optimization objectives of $\mathcal{S}$ and $\mathcal{T}$. In round $t$, given a noise vector $z \in Z$, we assume that after the training of $\mathcal{S}$ (line 7-11), ${\arg \max}_{i} \mathcal{S}_i(\mathcal{G}(z; \theta_g^{(t-1)}); \theta_s^{(t)}) = {\arg \max}_{i} \mathcal{T}_i(\mathcal{G}(z; \theta_g^{(t-1)}))$ and after the training of $\mathcal{G}$ (line 12-15), the inference confidence of the new synthetic examples $\mathcal{G}(Z)$ is higher on $\mathcal{S}$ and $\mathcal{T}$. We empirically justify these assumptions in Appendix~\ref{appendix:justify}.  Based on the assumptions above, we have Theorem~1 that shows $\mathcal{L}_{\mathcal{S}}$ can converge in our proposed algorithm. We also verify the convergence performance of our algorithm in the experiments. 

\noindent \textbf{Theorem 1.}  Given a noise vector $z \in Z$, let $f\left(\theta_{s}^{(t)}\right)=\mathrm{CE}\left(\mathcal{T}\left(\mathcal{G}(z; \theta_g^{(t)})\right), \mathcal{S}\left(\mathcal{G}(z; \theta_{g}^{(t)}); \theta_{s}^{(t)}\right)\right)$, where $t$ indexes the round in Algorithm~\ref{alg:code}, then $f(\theta_s^{(t)})$ can converge to a positive minimum.
\emph{(Proof in Appendix~\ref{proof:theorm1})}

\textbf{The cross entropy loss:}
In order to imitate the target model, the output of the substitute model needs to be as similar to the output of the target model as possible. The two models can be regarded as two distributions. The most straightforward way to enforce their similarity is to minimize the Kullback-Leibler divergence~\cite{kullback1951kld} between them. Here, we make use of cross entropy to transfer the generalization ability of the target model to the substitute model \cite{hinton2015distilling}. In our black-box setting, the KL-divergence loss and the cross entropy loss can be regarded as equivalent loss functions for stochastic optimization (see the appendix for more details).

\subsection{Black-box adversarial attacks}
Here, we illustrate how to launch adversarial attacks based on stolen substitute models. We start with the notations of adversarial attacks, their types, and their algorithms. 
Given an input $x \in X$, $\mathcal{T}(x)=y$, where $y$ is the inferenced label
of model $\mathcal{T}$. There are two types of data to $\mathcal{T}$: benign examples and adversarial examples, denoted by $\bar{x}$ and  $\widehat{x}$, respectively. Let $\bar{y}$ be the inference label of $\mathcal{T}$ for $\bar{x}$. Adversarial attacks aims to generate a visually indistinguishable example $\widehat{x} = \bar{x} + \epsilon$ to fool the target model, where $\epsilon$ is the perturbation of the normal example $\bar{x}$, and $\widehat{x}$ is called an adversarial example. There are two types of adversarial scenarios: \textit{untargeted} and \textit{targeted}, where the former misleads the target model to misclassify the adversarial examples, and the later leads the target model to a particular type of misclassification. In an untargeted attack, attackers try to shift the output of the target model over the adversarial example by  minimizing $\|\epsilon\|$ such that $\mathcal{T}\left(\widehat{x} \right) \neq \bar{y}$. In targeted attacks, adversarial attacks try to make the target model to classify the adversarial example to a particular label, i.e., $y_l$,
$\mathcal{T}\left( \widehat{x} \right) = y_{l}$.
Attackers take a benign example and then add perturbation according to the gradient of $\bar{x}$ in the target model by the function $\epsilon = P(\nabla_{\bar{x}}\mathcal{T}(\bar{x}))$, e.g., FGSM~\cite{Goodfellow:2015:FGSM}, where $P$ is the perturbation function over the gradient. Thus, for both targeted and untargeted attacks, generating $\epsilon$ for $\bar{x}$ requires knowledge of target model which can be either known or unknown to attackers, corresponding to white or black-box attacks, respectively. In white-box settings, attackers can directly access $\mathcal{T}$ to generate adversarial examples by adding $\epsilon$. However, in black-box attacks, as the attackers can't access the target model, a substitute model $\mathcal{S}$ that imitates the target model is required to generate adversarial examples. Thus, our model stealing process to obtain $\mathcal{S}$ is essential for black-box adversarial attacks.




%% file: sections/experiments.tex
\label{sec:exp}

In this section, we comprehensively evaluate the model stealing performance (the accuracy of the substitute model) on the challenging label-only scenario. In order to compare \alg with DaST~\cite{zhou2020dast} and DFME\footnote{For the evaluation of DFME, we have applied the original code from this paper with the recommended hyper-parameters. Our diligent attempts to optimize the hyper-parameters in our setting also do not have apparent improvement.}~\cite{truong2021data}, the state of the art data-free model stealing approaches, the evaluation is also extended to probability-only scenario. We also demonstrate that only a small number of queries are needed for \alg to learn substitute models and craft adversarial examples. Furthermore, we apply the learned substitute model as inputs to adversarial attacks that aim to fool the target models. 
To show the adversarial attacking performance, we evaluate the attack success rate under both targeted and untargeted scenarios, i.e., flipping the (un)targeted labels. 

\subsection{Experimental setup} 
\textbf{Dataset and Model Structure:} We evaluate our proposed method on three datasets: MNIST~\cite{lecun:1998:mnist}, Fashion-MNIST~\cite{xiao:2017:online:fashion} and CIFAR-10~\cite{krizhevsky:2009:cifar}. Note that for each dataset we use different network structures for the target and substitute models as the attackers lack of prior knowledge of the target structure. For MNIST, we utilize a lightweight CNN of four and three convolutional layers for the target and substitute models, respectively. Both  $\mathcal{S}$ and $\mathcal{T}$ structures of Fashion-MNIST are the same with MNIST. As for Cifar-10, the commonly adopted ResNet34 is used for $\mathcal{T}$ and we select a CNN of four convolutional layers for $\mathcal{S}$ (as DaST and DFME) in the experiments. 

\textbf{Attacking Scenarios and Methods: }In this paper, we focus on the
\textbf{label-only} scenario, where the attackers can access the output label only. In order to compare with baseline methods, \textbf{probability-only}\footnote{In the probability-only scenario, $\mathcal{T}(x)$ is probability vector as output.} scenario is also evaluated, where the accessible output is class probabilities. Here we also assume that the attacks in both scenarios are free to query $\mathcal{T}$ for unlimited times. To generate adversarial examples to fool the target model, we apply three popular attacking methods including \textbf{FGSM}~\cite{Goodfellow:2015:FGSM}, \textbf{BIM}~\cite{Kurakin:17:BIM}, and projected gradient descent (\textbf{PGD})~\cite{Madry:18:PGD}. 

\textbf{Evaluation Criteria: }The goal of our  data-free model stealing is to achieve high classification \textbf{accuracy} on substitute model. The application of adversarial attack aims at misleading $\mathcal{T}$ by both untargeted and targeted attacks.  We denote them as \alg-U and \alg-T respectively. In the experiments, we use the second class as target for all three datasets for targeted attacks.
The criteria to evaluate both types of attacks is the attack success rate (\textbf{ASR}), defined by $n_{suc}/n_{all}$, where $n_{all}$ is the number of generated adversarial examples to fool $\mathcal{T}$, and $n_{suc}$ is the number of successful attempts. For evaluating the attacking efficiency, we use the number of queries to reach the maximum accuracy of the substitute model as the criteria.

\subsection{Model Stealing Performance}

\begin{figure}[h]
	\centering
	{
	\begin{subfigure}{0.49\linewidth}
	    \includegraphics[width=0.9\textwidth]{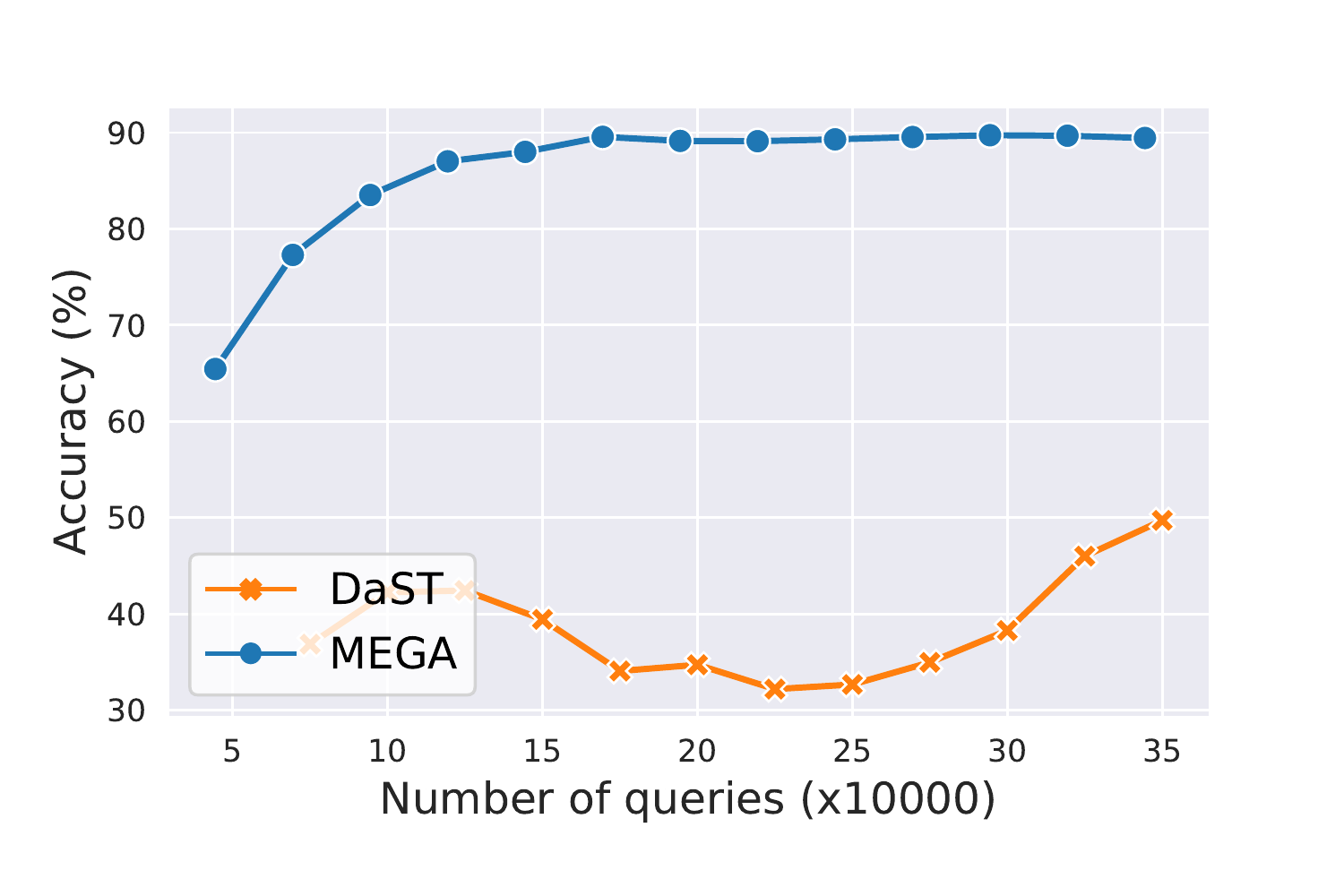}
        \caption{MNIST, label-only}
	    \label{subfig:mnist_l_only}
	\end{subfigure}
	\begin{subfigure}{0.49\linewidth}
    \includegraphics[width=0.9\textwidth]{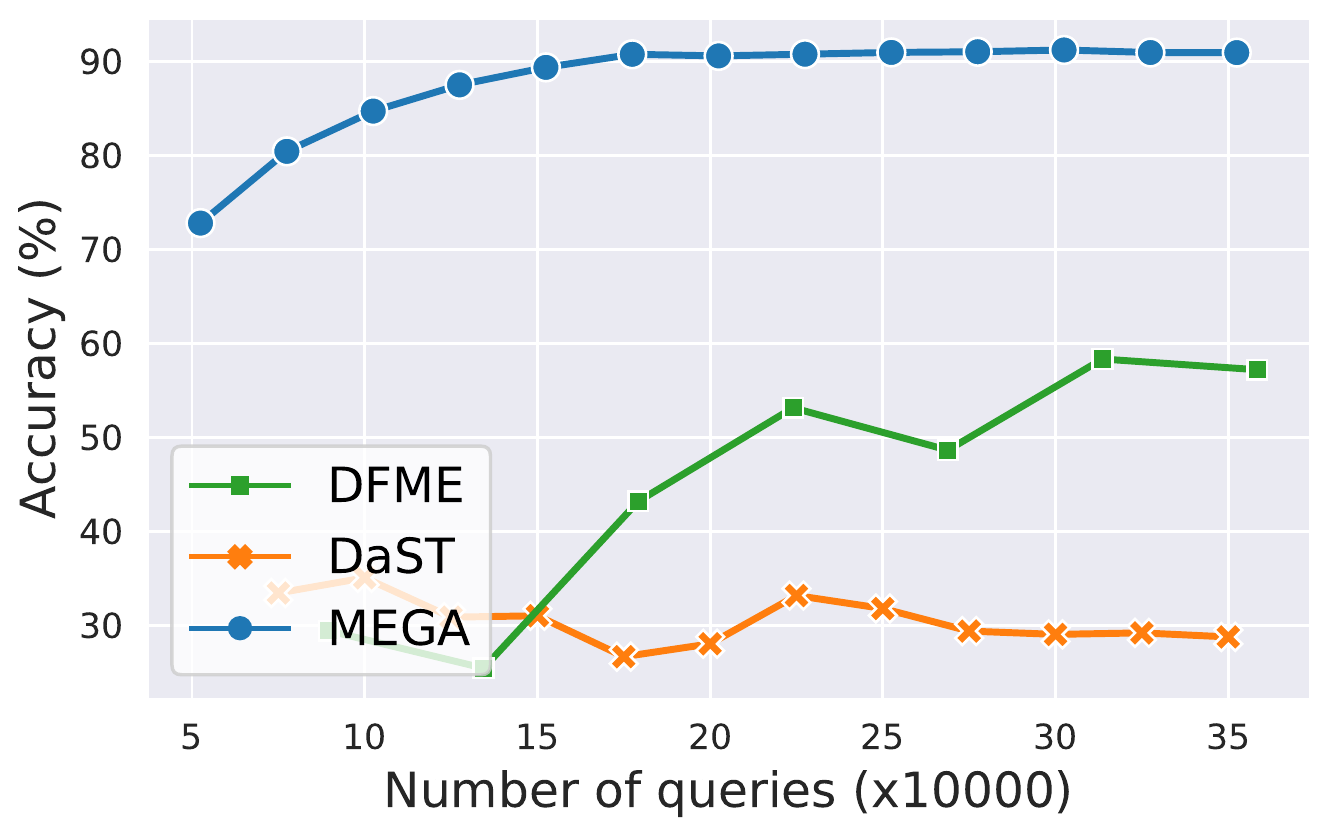}
    \caption{MNIST, probability-only}
    \label{subfig:mnist_p_only}
\end{subfigure}
	\hfill
	\begin{subfigure}{0.49\linewidth}
	    \includegraphics[width=0.9\textwidth]{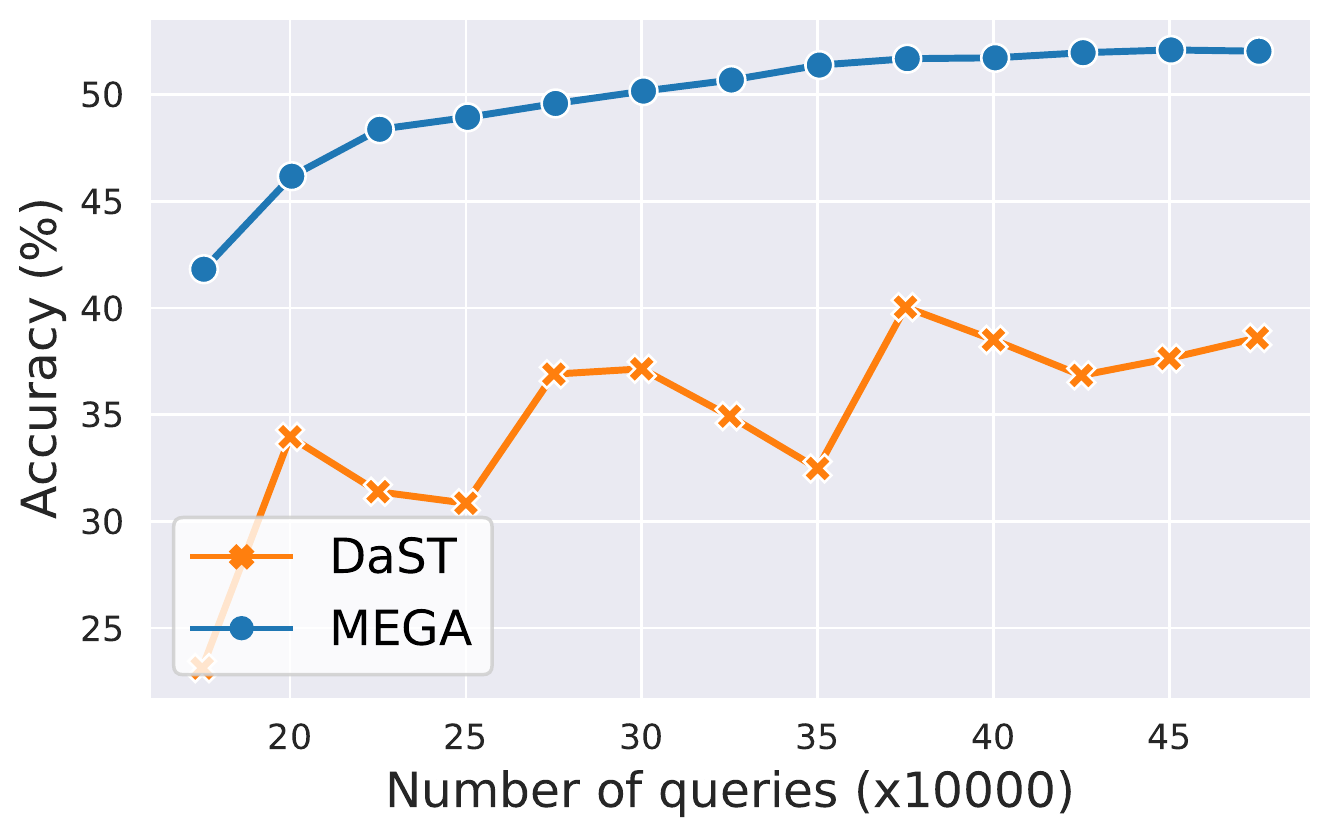}
        \caption{Fashion-MNIST, label-only}
	    \label{subfig:fashion_mnist_l_only}
	\end{subfigure}
    \begin{subfigure}{0.49\linewidth}
	    \includegraphics[width=0.9\textwidth]{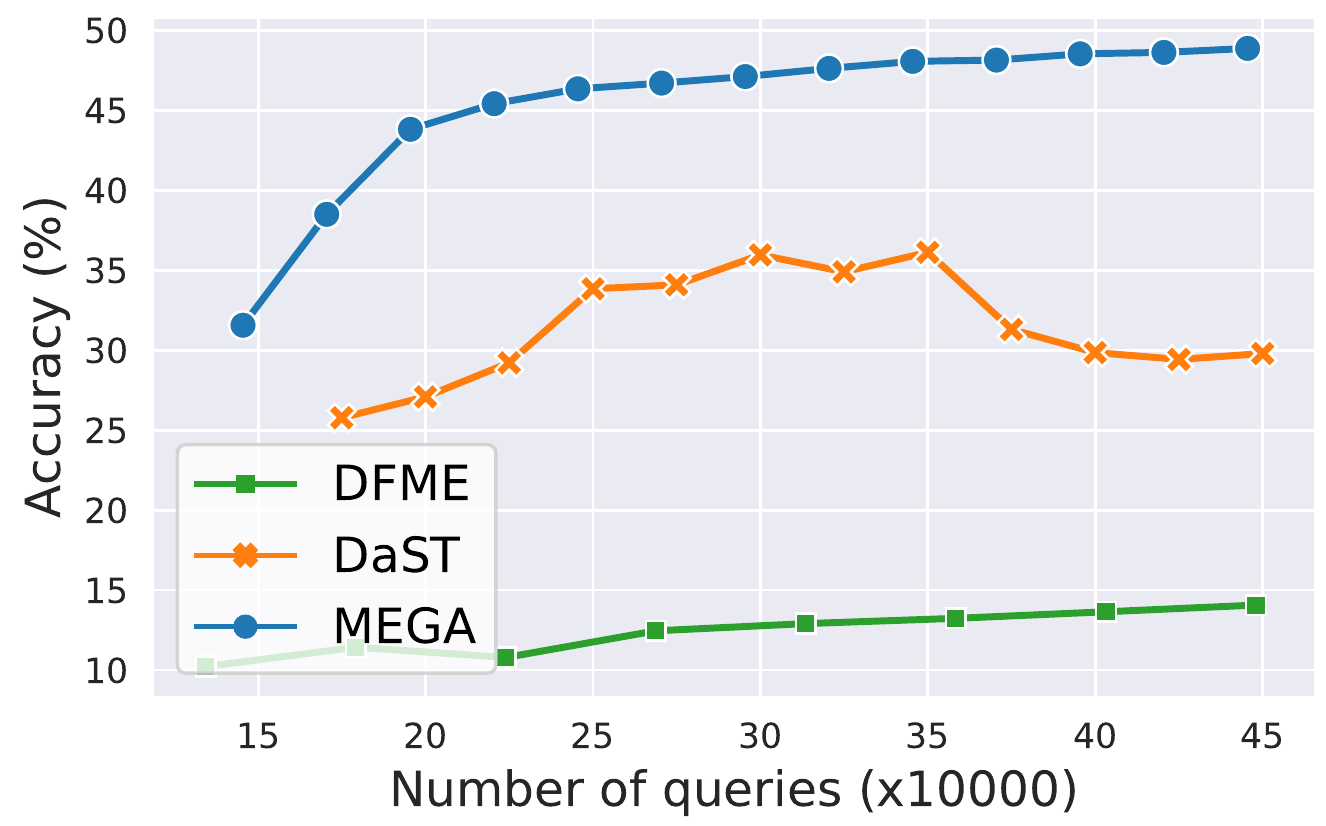}
	    \caption{Fashion-MNIST, probability-only}
	    \label{subfig:fashion_mnist_p_only}
	\end{subfigure}
	}
	\caption{The accuracy of the substitute models.}
	\label{fig:ms_acc}
\end{figure}

\textbf{Model Stealing Accuracy:} Here we evaluate the accuracy of model stealing results compared to baseline model of DaST and DFME on both label-only and probability-only scenarios in Table~\ref{tab:ms_acc}. 
Table~\ref{tab:ms_acc} first shows the accuracy of the target model, which is essentially the accuracy upper bound of the substitute model. It should be noted that DFME is only compared on probability-only scenario since DFME algorithm requires necessary additional information than inference labels. As DaST, DFME and \alg apply different updating strategies and batch sizes, it is hard to uniform them by training iterations or rounds. For a fair comparison, we use the number of queries as the horizontal axis\footnote{Here we limit the number of queries for plotting according to the convergence of \alg. Actually the training processes of DaST and DFME are far longer however its modeling stealing accuracy stays fluctuating and way below \alg. The entire results are provided in the appendix.}. We further note that DaST and DFME generate synthetic data per training iteration for generator or substitute model, whereas \alg updates the entire model via multiple training iterations per set of synthetic data.
\begin{figure}[h]
	\centering
	{
	\subfloat[Probability-only:MNIST]{
	    \label{subfig:ms_loss_codef}
	    \includegraphics[width=0.4\textwidth]{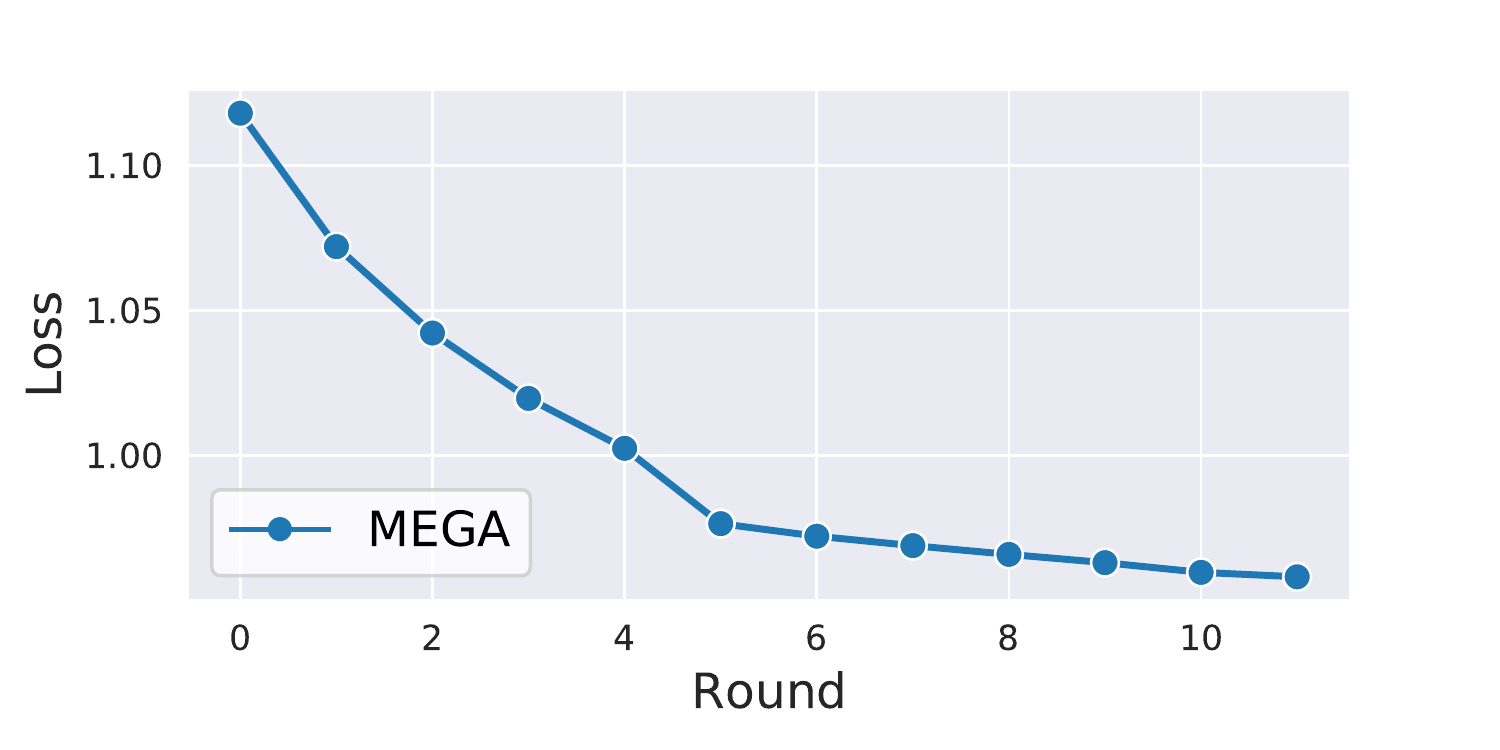} }
\hfill
    \subfloat[Label-only:MNIST]{
	    \label{subfig:ms_loss_dast}
	    \includegraphics[width=0.4\textwidth]{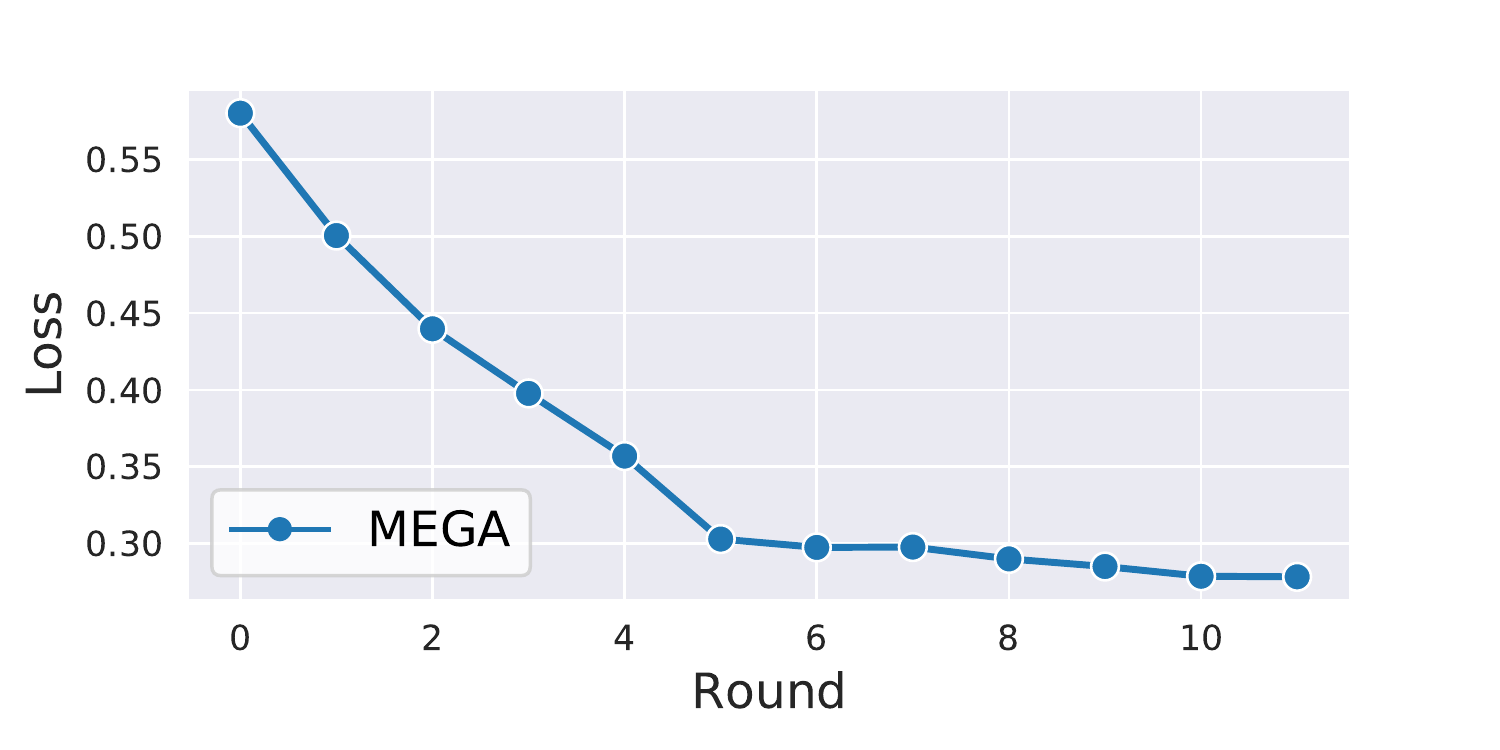}}
	}
	\caption{Convergence of \alg over training rounds: loss of the substitute model}
	\label{fig:ms_loss}
\end{figure}

\begin{table}
\renewcommand\arraystretch{1.25}
\centering
\caption{Comparison of the accuracy of the substitute models.}
\label{tab:ms_acc}
\resizebox{1\columnwidth}{!}{%
\begin{tabular}{c|c|cc|ccc} 
\toprule
& Target  & \multicolumn{2}{c|}{\textbf{Label-only}} & \multicolumn{3}{c}{\textbf{Probability-only}} \\
\textbf{Dataset} & Accuracy & DaST & \alg & DFME & DaST & \alg\\ 
\midrule
\textit{MNIST}  & 99.12 &         68.69      &        \textbf{89.43}            &  63.14 & 66.11 & \textbf{90.93}  \\ 
\textit{Fashion-MNIST}  & 91.15  &        33.07       &     \textbf{52.04}             & 15.90 &   29.74 & \textbf{48.89}        \\           

\bottomrule
\end{tabular}}
\end{table}

From Table.~\ref{tab:ms_acc}, it is clear that the substitute model accuracy of \alg is higher than DaST and DFME with a significant gap. For label-only scenarios, the accuracy of substitute of \alg on MINST dataset even approaches 90\%. It strongly demonstrates the effectiveness of \alg in stealing the target model in a data-free manner. On the other hand, the performance on model stealing on Fashion-MNIST shows a drop owing to the increasing difficulty of classification task. 
Even without high stealing accuracy on Fahsion-MNIST, it is interesting that it can still bring high adversarial attack success rate (see results later). For probability-only scenarios, DaST outperfoms DFME slightly but we cannot see very clear differences over rounds. Moreover, from Figure~(\ref{subfig:mnist_l_only}) to (\ref{subfig:fashion_mnist_p_only}), \alg can quickly reach high accuracy with a much smaller number of queries than DaST and DFME.  Another observation worth mentioning is that \alg converges to the maximum accuracy steadily in both label-only and probability-only scenarios, whereas DaST apparently oscillates in both.
We also provide the long run training process of DaST and DFME in the appendix to verify our statement on their model stealing performance.

\textbf{Convergence Process: }For DaST, we can see from Figure~(\ref{fig:ms_acc}) that the accuracy fluctuates and doesn't stop at a good local optima along the entire training procedure of 350,000 queries for (\ref{subfig:mnist_p_only})(\ref{subfig:mnist_l_only}) (450,000, 500,000 for (\ref{subfig:fashion_mnist_p_only})(\ref{subfig:fashion_mnist_l_only}) respectively). The accuracy of substitute model does not show an increasing trend in Figure~(\ref{subfig:mnist_p_only})(\ref{subfig:fashion_mnist_p_only}). Nevertheless, DaST saves the substitute model each iteration, and choose the one of highest accuracy to be the final result. 
We argue the it is infeasible to select the best model for DaST because attackers do not have real data to evaluate their saved models. Another issue is that what stopping criteria shall be used for terminating the training especially when the accuracy does not converge stably, e.g., shown in Figure.~(\ref{subfig:fashion_mnist_p_only}).
Further, once the model stealing process ends, it requires additional resources and efforts to store and to select the best substitute model. This further leads to the unstable training process of DaST. For DFME, the accuracy also oscillate a lot because it also applies competitive training. Comparatively, the convergence process of \alg is almost monotonic. The accuracy increases smoothly during the whole training phrase, and converges to local optima at around 200,000 queries for MNIST and 400,000 for Fashion-MNIST, respectively. The loss of substitute models shown in Figure~\ref{fig:ms_loss} stays consistent with our theoretical analysis on convergence guarantee in Section~\ref{subsec:analysis}.

There are multiple reasons behind such differences in convergence between DaST, DFME and \alg. First, \alg generates a set of synthetic data to update the generator and the substitute model via multiple training iterations, aiming to find the local optima for each set. On the contrary, DaST and DFME generates a new set of examples in each iteration to update the substitute model and the generator. As a result, each set is  used to guide only one training iteration. However, such one-iteration update provides limited guidance to the training, even worse, the knowledge learned by one update can be dominated by random factors of a new set of synthetic data. Second, the training process in DaST and DFME are similar to GANs~\cite{Cha21GAN, Jeong21Training} and inherit the limitations of GANs. 
Under GANs, real examples are applied to guide the training. However, real examples are limited in a small region of feature space, but for synthetic noisy images, each new set generated from different seeds can be totally different from the others, which brings extra instability especially in the case of synthetic examples from a large random generating space. Thus, in \alg, we only use a fixed set of random seeds and fully exploit them in multiple rounds. Thirdly, the objective function of the generator of DaST and DFME are trained in a competition game with $\mathcal{S}$, but for our confidence-based objective, the training procedure is a collaborative game among them. The optimization objectives of $\mathcal{S}$ and $\mathcal{G}$ are complimentary rather than contradictory, and thus enhances stability.


\subsection{Model Stealing Efficiency}


\begin{table}
\renewcommand\arraystretch{1.25}
\centering
\caption{The number of queries ($\times 10000$) to achieve the maximum substitute model accuracy for DaST, DFME and \alg.}
\label{tab:queries}
\resizebox{1\columnwidth}{!}{%
\begin{tabular}{c|cc|ccc} 
\toprule
 & \multicolumn{2}{c|}{\textbf{Label-only}} & \multicolumn{3}{c}{\textbf{Probability-only}} \\
\textbf{Dataset} & DaST & \alg & DFME & DaST & \alg\\ 
\midrule
 \textit{MNIST}  &          2150     & \underline{\textbf{34}}      & 72 &           590          &           \underline{\textbf{35}}                \\ 
 \bottomrule
\textit{Fashion-MNIST}  & 75 &     \underline{\textbf{48}}    & 116 &           90          &           \underline{\textbf{45}}              \\           
\bottomrule
\textit{Cifar-10}   & 30 &     \underline{\textbf{20}}   & 269 &           100          &           \underline{\textbf{17}}             \\
\bottomrule
\end{tabular}
}
\end{table}

Another performance superiority of \alg is model stealing efficiency, compared to other data-free model stealing.
We zoom into the metrics to show efficiency of the proposed model stealing process: the number of queries needed to achieve the maximum accuracy of a substitute model.
The results of number of queries for training are summarized in Table~\ref{tab:queries}.  One can see that \alg requires 37\%--95\% fewer queries to train the substitute models than DaST and DFME. 
As \alg has light-weight networks, a smaller number of query-label pairs is needed and the required training time is thus lower. Another reason lies on the convergence speed. As the training of substitute model for both DFME and DaST fluctuates a lot, it is hard to find the terminating point of training process so that more iterations are needed to obtain better $\mathcal{S}$. 
In a nutshell, \alg not only requires a lower number of synthetic query examples but also lower training time than DaST and DFME.

\subsection{Adversarial Attacks based on Stolen Models}

\begin{table*}
\renewcommand\arraystretch{1.25}
\centering
\caption{The attack success rate on three datasets for \alg and DaST under untargeted and targeted attacks of probability-only and label-only scenarios.  }
\label{tab:asr}
\resizebox{2\columnwidth}{!}{%
\begin{tabular}{c|c|cccc|cccccc} 
\toprule
& & \multicolumn{4}{c|}{\textbf{Label-only}} & \multicolumn{6}{c}{\textbf{Probability-only}} \\
\textbf{Dataset} & \textbf{Attack} &  DaST-U & \alg-U  & DaST-T & \alg-T  & DFME-U & DaST-U & \alg-U & DFME-T & DaST-T &  \alg-T  \\ 
\midrule
 & \textit{FGSM}  &   20.22  & \underline{\textbf{53.72}} &  6.32    &\underline{\textbf{16.47}}          & 30.08 &          24.65           &           \underline{\textbf{52.00}}                           &  8.01 &       6.33            &          \underline{\textbf{16.04}}                   \\ 
MNIST & \textit{BIM}  & 27.83 &     \underline{\textbf{65.26}} &  2.07     & \underline{\textbf{18.94}}       & 39.05 &           27.93          &           \underline{\textbf{61.76}}                      &    12.35 &      2.49           &           \underline{\textbf{15.42}}                        \\ 
& \textit{PGD}  &   27.73   & \underline{\textbf{65.19}}  & 2.13  &  \underline{\textbf{18.97}}          & 38.38  &          28.42           &             \underline{\textbf{60.92}}                  &    12.12 &     2.78            &          \underline{\textbf{15.33}}                    \\ 

\bottomrule
 & \textit{FGSM}  &   \underline{\textbf{88.77}}  &  86.55   &          9.81        &  \underline{\textbf{17.61}}                &  49.33  &              87.76     &  \underline{\textbf{89.08}}                &     0.54 &    2.10 & \underline{\textbf{30.74}}               \\ 
Fashion-MNIST & \textit{BIM}  &          85.10  & \underline{\textbf{87.10 }}   & 23.46       &  \underline{\textbf{28.46}}     &  49.22 &  88.50         & \underline{\textbf{90.42}}            & 0.31 &  6.18&        \underline{\textbf{37.28}}                 \\ 
& \textit{PGD}  &85.60    &   \underline{\textbf{88.69}}    & 23.55      & \underline{\textbf{28.63}}    &     51.21 &     89.93         &             \underline{\textbf{92.07}}              &    0.27 &     5.71            &          \underline{\textbf{37.75}}                \\ 

\bottomrule
 & \textit{FGSM}  &  54.62  & \underline{\textbf{57.99}}  & 0.16   &\underline{\textbf{0.22}}               &      53.69 &    56.06           &           \underline{\textbf{58.60}}                    &     0.17  &   0.20  & \underline{\textbf{0.22}}               \\ 
Cifar-10 & \textit{BIM}   & 56.25 &     \underline{\textbf{60.02}}    &  0.17   & \underline{\textbf{0.20}}       &     47.37 &      58.54          &           \underline{\textbf{60.18}}                     &    0.23 &      0.27           &           0.27                      \\ 
& \textit{PGD}    &   56.15   & \underline{\textbf{59.50}}  & 0.12 &  \underline{\textbf{0.19}}      &      47.19 &    58.52         &             \underline{\textbf{59.95}}             & 0.22          &         \underline{\textbf{0.29}}   & 0.25                   \\ 
\bottomrule
\end{tabular}
}
\end{table*}

\begin{table}
\renewcommand\arraystretch{1.25}
\centering
\caption{The attack success rate on MNIST using unlabeled real data to steal the target model.}
\label{tab:real_asr}
\resizebox{0.8\columnwidth}{!}{%
\begin{tabular}{c|cc|cc} 
\toprule
& \multicolumn{2}{c|}{\textbf{Label-only}} & \multicolumn{2}{c}{\textbf{Probability-only}} \\
\textbf{Attack} & Real-U & Real-T & Real-U & Real-T \\ 
\midrule
\textit{FGSM}        &         61.80   & 15.68              &         63.95   &  16.38 \\ 
\textit{BIM}          &      80.24  &     30.06      &          79.02           & 29.95      \\ 
\textit{PGD}         &          80.46   & 30.20        &      78.77     &   29.48           \\ 

\bottomrule
\end{tabular}
}
\end{table}

Here we evaluate the attack success rate in both \textit{targeted} and \textit{untargeted} adversarial under label-only and probability-only scenarios on three datasets. The experimental results by ASR are summarized in Table~\ref{tab:asr}. Cifar-10 dataset is chosen here to show the results of the data-free attack among more complex network and task. 
Prior to discussing the ASR of \alg, we first show the ASR achieved by real data to steal the target model for MNIST dataset in Table~\ref{tab:real_asr}.  Here we use real data to query the target model so as to train the substitute model. Then we apply the same attack methods for generating adversarial examples. It can be regarded as the upper-bound of ASR for our data-free settings. Overall, the attack success rate of \alg reaches around 85\% of the real data case for untargeted attacks and worse results for targeted attacks depending on the attack methods. From Table~\ref{tab:asr}, it is clear to see that ASR of \alg significantly outperforms DaST and DFME from two to fifteen times high, but DaST shares the similar performance as DFME in general. Comparing untargeted and targeted attacks, all of DaST, DFME and \alg show higher ASR for untargeted attack. This can be explained by that a untargeted attack only need classification to shift out of the correct label while the a targeted attack requires an elaborate $\epsilon$ to lead to the specific misclassification outcome. As a result, more significant ASR improvement resulted from \alg can be observed for the challenge targeted cases than the untargeted ones. Also, the result that \alg achieves about 55\% high of real data scenario is as expected, because more elaborate $\epsilon$ is required but too demanding for synthetic data.

When it comes to differences between label-only and probability-only, higher attack success rates are achieved under the probability-only scenarios. This can be explained by that in label-only scenarios, attackers train the substitute model by pairs of synthetic data $X$ and querying output labels. Specifically, the learning of the substitute model is to imitate the output label of $\mathcal{T}$ by cross entropy loss. Thus, when outputting the same label (class element of maximum probability), it does not provide information for training. On the contrary, the feedback the probability vector of all classes provides much more supervision to  the substitute model. When $\mathcal{S}$ and $\mathcal{T}$ output the same label of maximum probability, $\mathcal{S}$ continues to learn until all of the probabilities become similar. It is in line with our previous statement that probability-only scenarios are more informative, and achieve stronger attacks. 

Let us zoom into the differences among datasets. In general, \alg greatly outperforms DaST and DFME in all three datasets. Compared to MNIST, the attack success rate of Fashion-MNIST is higher although the accuracy of $\mathcal{S}$ in Figure.~\ref{fig:ms_acc} is lower. It is because that the classification task of MNIST is very easy, adding perturbation helps more for attacking when the task is not that easy. However, despite the fact that Cifar-10 task is more challenging than Fashion-MNIST, the attack success rate on Cifar-10 appears to be lower. The reason of such results lies on the fact of low accuracy of the substitute model. Training on Cifar-10 even with real data and deep networks can be difficult~\cite{ccalik2018cifar,thakkar2018batch}. Without surprise, shallow substitute models trained by synthetic data thus reach even lower accuracy. This observation unfortunately implies the limitation of data-free adversarial attacks in complex learning tasks that need deeper network structures with massive training data. 

%% file: sections/appendix.tex
\section{Proof of Theorem~1}
\label{proof:theorm1}

In this section, we prove Theorem~1 which shows that the loss of the substitute model $\mathcal{S}$ can converge during the training. In order to simplify the proof of Theorem~1, we firstly derive Lemma~1 according to the assumptions in Section~\ref{subsec:conv}.

\subsection{Notations and lemmas}

\label{sec:lemma1}

For simplicity, we use $\mathcal{T}\left(z; \theta_g^{(t)}\right)$ to denote $\mathcal{T}\left(\mathcal{G}(z; \theta_g^{(t)})\right)$ and $\mathcal{S}\left(z; \theta_{g}^{(t)}, \theta_{s}^{(t)}\right)$ to denote $\mathcal{S}\left(\mathcal{G}(z; \theta_{g}^{(t)}); \theta_{s}^{(t)}\right)$. Then we can derive the following lemma. The lemma shows that training $\mathcal{G}$ using $\mathcal{L}_{\mathcal{G}}$ also indirectly decreases the loss of $\mathcal{S}$.

\noindent \textbf{Lemma 1.} \emph{After the training of $\mathcal{G}$ (line 11-14, Algorithm~\ref{alg:code}) in round $t$, given $z \in Z$, we have $\mathrm{CE}(\mathcal{T}(z; \theta_g^{(t)}), \mathcal{S}(z; \theta_{g}^{(t)}, \theta_{s}^{(t)})) \leq \mathrm{CE}(\mathcal{T}(z; \theta_g^{(t-1)}), \mathcal{S}(z; \theta_{g}^{(t-1)}, \theta_{s}^{(t)}))$.}

\emph{Proof.} After training $\mathcal{G}$ we have 

\begin{equation*}
    \mathcal{S}_{i^{*}}(z, \theta_g^{(t)}, \theta_s^{(t)}) \geq \mathcal{S}_{i^{*}}(z, \theta_g^{(t-1)}, \theta_s^{(t)}),
\end{equation*}
\begin{equation*}
    \mathcal{T}_{i^{*}}(z, \theta_g^{(t)}, \theta_s^{(t)}) \geq \mathcal{T}_{i^{*}}(z, \theta_g^{(t-1)}, \theta_s^{(t)}),
\end{equation*}
where $i^{*} = {\arg \max}_i \mathcal{S}_i(\mathcal{G}(z; \theta_g^{(t-1)}); \theta_s^{(t)}) = {\arg \max}_i \mathcal{T}_i(\mathcal{G}(z; \theta_g^{(t-1)}))$. Then we have

\begin{equation*}
\begin{aligned}
    \mathrm{CE}(\mathcal{T}(z; \theta_g^{(t)}),& \mathcal{S}(z; \theta_{g}^{(t)}, \theta_{s}^{(t)})) \\
    &= - \sum_{i=1}^{N} \mathcal{T}_{i}(z; \theta_g^{(t)}) \log \mathcal{S}_{i}(z; \theta_g^{(t)}, \theta_s^{(t)}) \\
    &\leq - \sum_{i=1}^{N} \mathcal{T}_{i}(z; \theta_g^{(t-1)}) \log \mathcal{S}_{i}(z; \theta_g^{(t)}, \theta_s^{(t)}) \\
    &\leq - \sum_{i=1}^{N} \mathcal{T}_{i}(z; \theta_g^{(t-1)}) \log \mathcal{S}_{i}(z; \theta_g^{(t-1)}, \theta_s^{(t)}) \\
    & = \mathrm{CE}(\mathcal{T}(z; \theta_g^{(t-1)}), \mathcal{S}(z; \theta_{g}^{(t-1)}, \theta_{s}^{(t)}))
\end{aligned}
\end{equation*}

\subsection{Completing the proof}
\noindent \textbf{Theorem 1.} \emph{Given $z \in Z$. Let $f\left(\theta_{s}^{(t)}\right)=\mathrm{CE}\left(\mathcal{T}\left(\mathcal{G}(z; \theta_g^{(t)})\right), \mathcal{S}\left(\mathcal{G}(z; \theta_{g}^{(t)}); \theta_{s}^{(t)}\right)\right)$. Training the substitute model by Algorithm~1, we have $\lim_{t \to \infty} f(\theta_s^{(t)}) = \epsilon^{*}$, where $\epsilon^{*} \geq 0$.}

\emph{Proof.} We can simplify $f\left(\theta_{s}^{(t)}\right)$ as 

\begin{equation*}
f\left(\theta_{s}^{(t)}\right)=\mathrm{CE}\left(\mathcal{T}\left(z; \theta^{(t)}\right), \mathcal{S}\left(z; \theta_{g}^{(t)}, \theta_{s}^{(t)}\right)\right),
\end{equation*}
where $t$ is used to index the training rounds. Using Lemma~1, We have 

\begin{equation*}
f\left(\theta_{s}^{(t+1)}\right) \leq \mathrm{CE}\left(\mathcal{T}\left(z; \theta g^{(t)}\right), S\left(z; \theta_{g}^{(t)}, \theta_{s}^{(t+1)}\right)\right).
\end{equation*}
Since the cross entropy loss is the loss function of $\mathcal{S}$, we have

\begin{equation*}
\begin{aligned}
f\left(\theta_{s}^{(t+1)}\right) &\leq \mathrm{CE}\left(\mathcal{T}\left(z, \theta g^{(t)}\right), \mathcal{S}\left(z, \theta_{g}^{(t)}, \theta_{s}^{(t+1)}\right)\right)\\ 
&\leq \mathrm{CE}\left(T\left(z, \theta g^{(t)}\right), S\left(z, \theta_{g}^{(t)}, \theta_{s}^{(t)}\right)\right)\\
&=f\left(\theta_{s}^{(t)}\right) 
\end{aligned}
\end{equation*}

Therefore, we know that $f\left(\theta_{s}\right)$ is monotone decreasing during the training. $f\left(\theta_{s}\right) = 0$ if and only if $\mathcal{T}\left(z; \theta^{(t)}\right) = \mathcal{S}\left(z; \theta_{g}^{(t)}, \theta_{s}^{(t)}\right)$. Otherwise $f\left(\theta_{s}\right) > 0$. Since $f\left(\theta_{s}\right) \geq 0$, it will converge. However the outputs of $\mathcal{S}$ and $\mathcal{T}$ usually won't be exactly the same. Then the convergence can be formally represented as $\lim_{t \to \infty} f(\theta_s^{(t)}) = \epsilon^{*}$, where $\epsilon^{*}$ is a constant and $\epsilon^{*} \geq 0$.

In this section, the convergence of \alg is analyzed based on the two assumptions shown in Section~\ref{subsec:analysis}. The two assumptions are empirically justified in Appendix~\ref{appendix:justify}. It does not seem possible to rigorously derive Theorem~1 without these two assumptions.

\section{Justifying assumptions}
\label{appendix:justify}

Here, we empirically justify the assumptions proposed in the methodology part. These assumptions are made for convergence analysis.

\subsection{Inference confidence}
\begin{figure}[h]
	\centering
	{
	\begin{subfigure}{0.49\linewidth}
	    \includegraphics[width=0.9\textwidth]{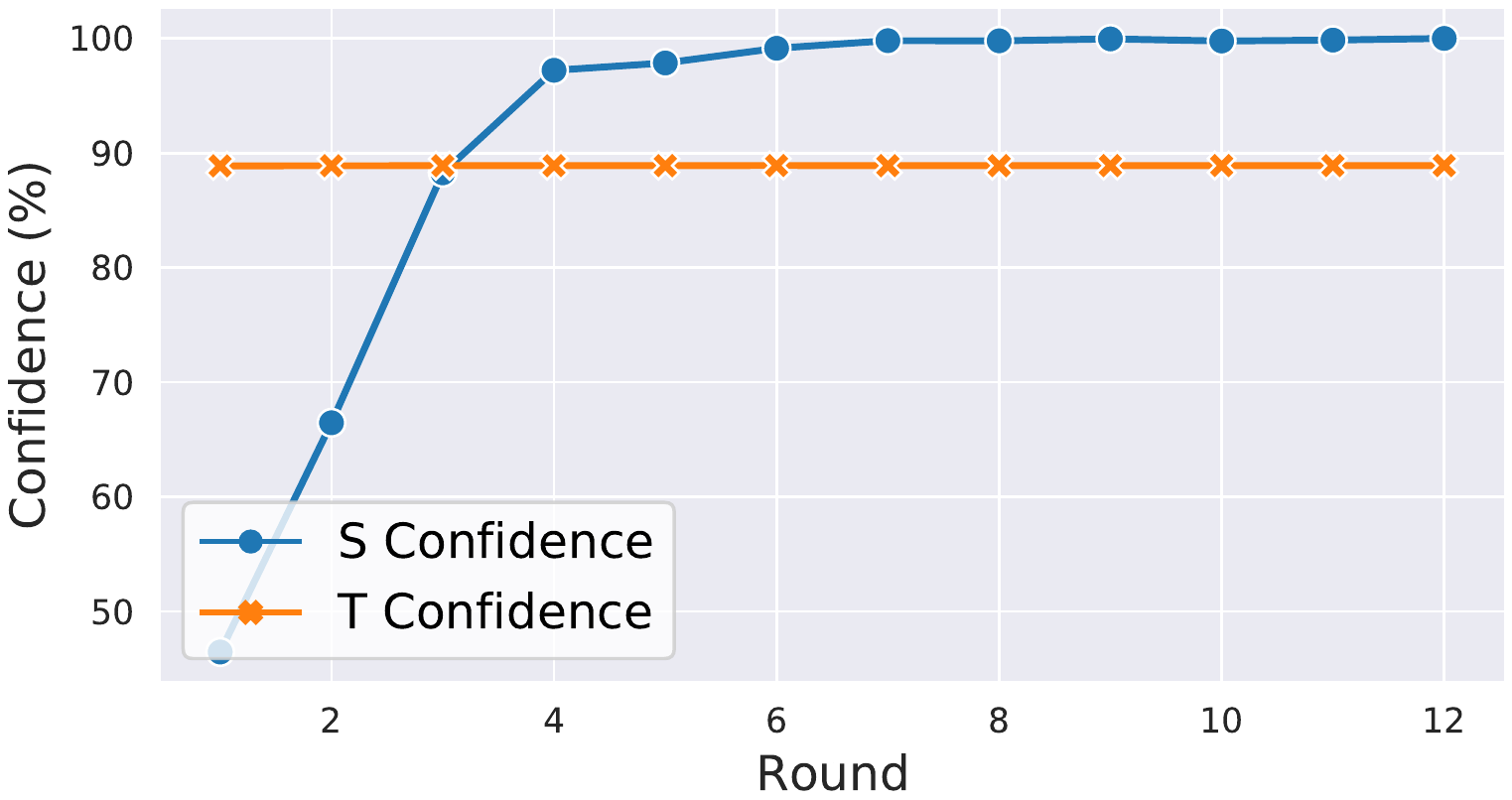}
        \caption{$z_1$:Label-only}
	    \label{subfig:mnist_l_only}
	\end{subfigure}
	\begin{subfigure}{0.49\linewidth}
    \includegraphics[width=0.9\textwidth]{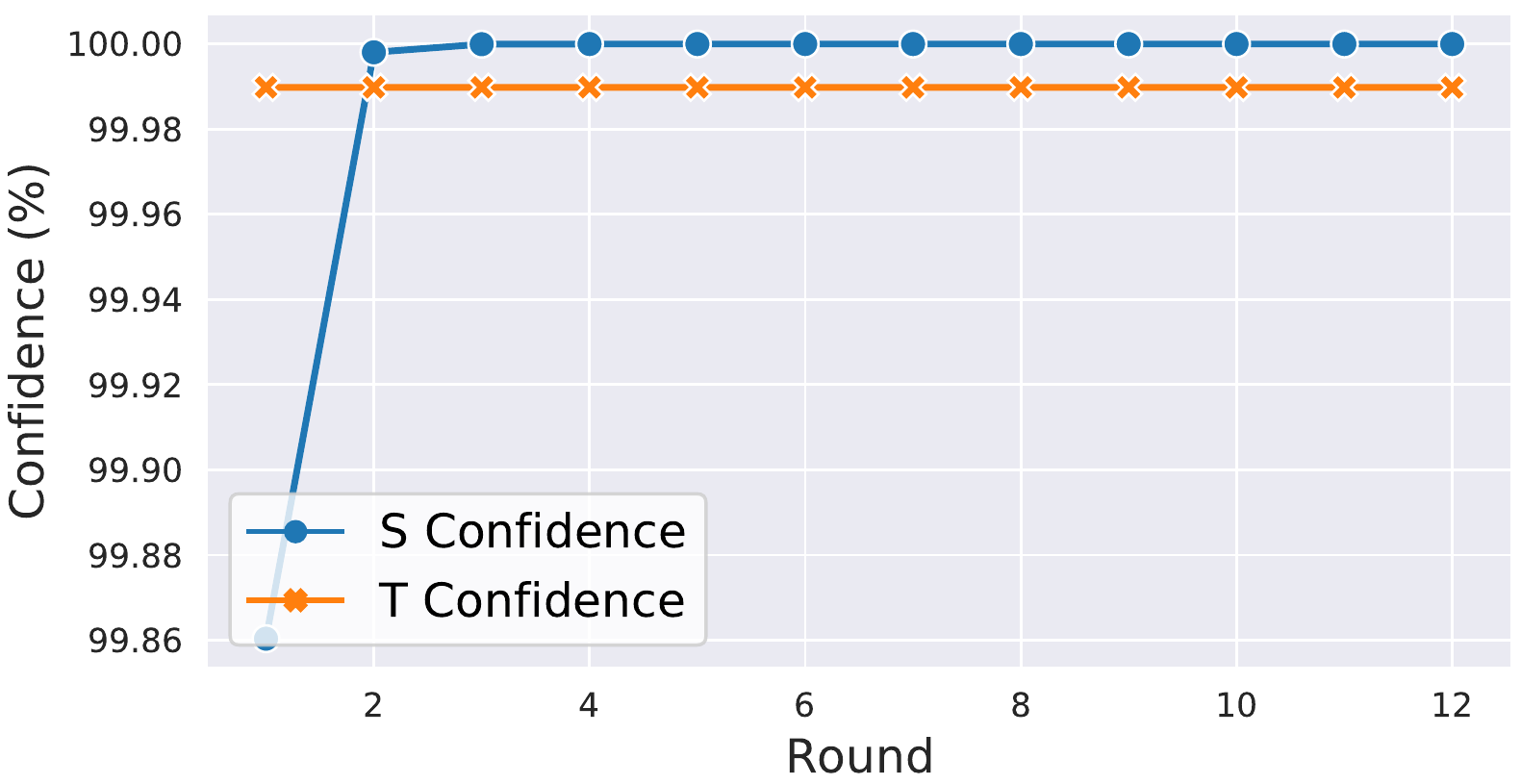}
    \caption{$z_2$:Label-only}
    \label{subfig:mnist_p_only}
\end{subfigure}
	\hfill
	\begin{subfigure}{0.49\linewidth}
	    \includegraphics[width=0.9\textwidth]{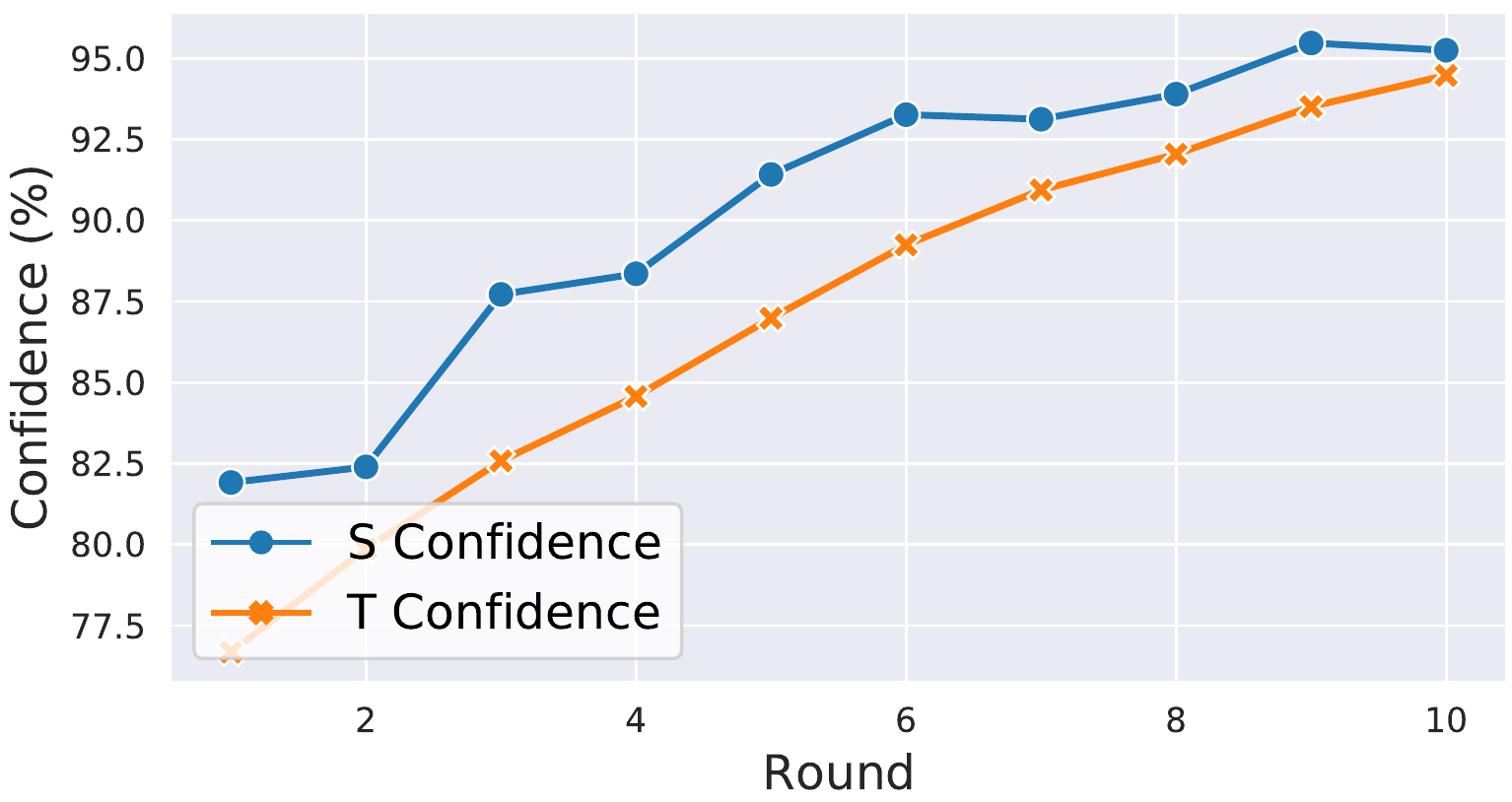}
        \caption{$z_3$:Probability-only}
	    \label{subfig:fashion_mnist_l_only}
	\end{subfigure}
    \begin{subfigure}{0.49\linewidth}
	    \includegraphics[width=0.9\textwidth]{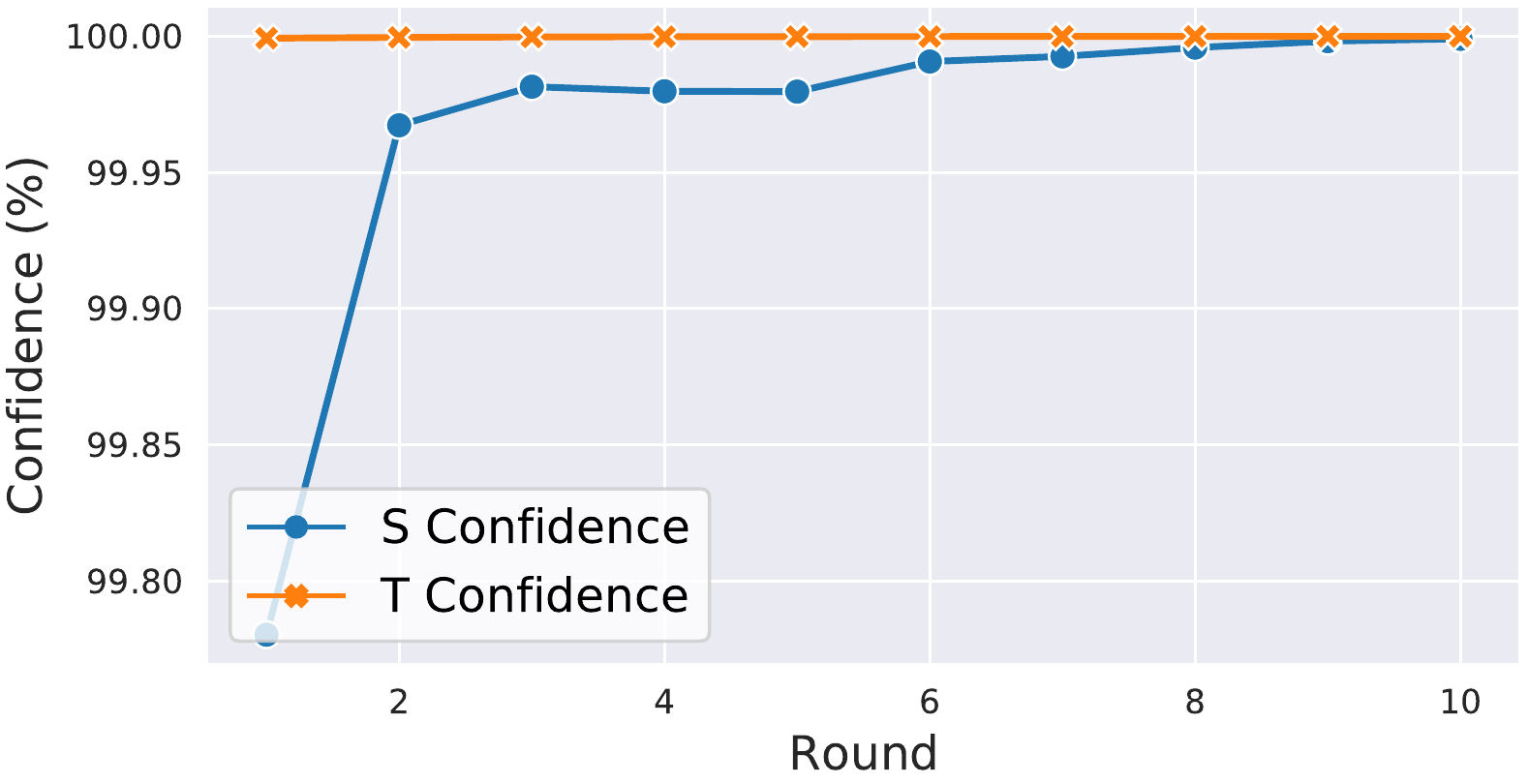}
	    \caption{$z_4$:Probability-only}
	    \label{subfig:fashion_mnist_p_only}
	\end{subfigure}
	}
	\caption{The inference confidence on different noise seeds in MNIST.}
	\label{supp:lconf}
\end{figure}

In Section~\ref{subsec:analysis}, we assume that after the training of $\mathcal{G}$ (algorithm line 12-15), the inference confidence of the new synthetic examples $\mathcal{G}(Z)$ is higher on $\mathcal{S}$ and $\mathcal{T}$. To justify this assumption, we show the inference confidence in different round in Figure~\ref{supp:lconf}. The confidence of four noise seeds are illustrated and the seeds are applied in both label-only and probability-only scenarios. In probability-only scenario, the inference confidence of $\mathcal{S}$ and $\mathcal{T}$ are similar because $\mathcal{S}$ can get the inference probability information from $\mathcal{T}$.

\subsection{Prediction similarity}
In Section~\ref{subsec:analysis}, we also assume that given a noise vector $z \in Z$, we assume that after the training of $\mathcal{S}$ (algorithm line 7-11), ${\arg \max}_{i} \mathcal{S}_i(\mathcal{G}(z; \theta_g^{(t-1)}); \theta_s^{(t)}) = {\arg \max}_{i} \mathcal{T}_i(\mathcal{G}(z; \theta_g^{(t-1)}))$.

\begin{figure}[h]
	\centering
	{
	\subfloat[Label-only:MNIST]{
	    \label{subfig:ms_loss_codef}
	    \includegraphics[width=0.35\textwidth]{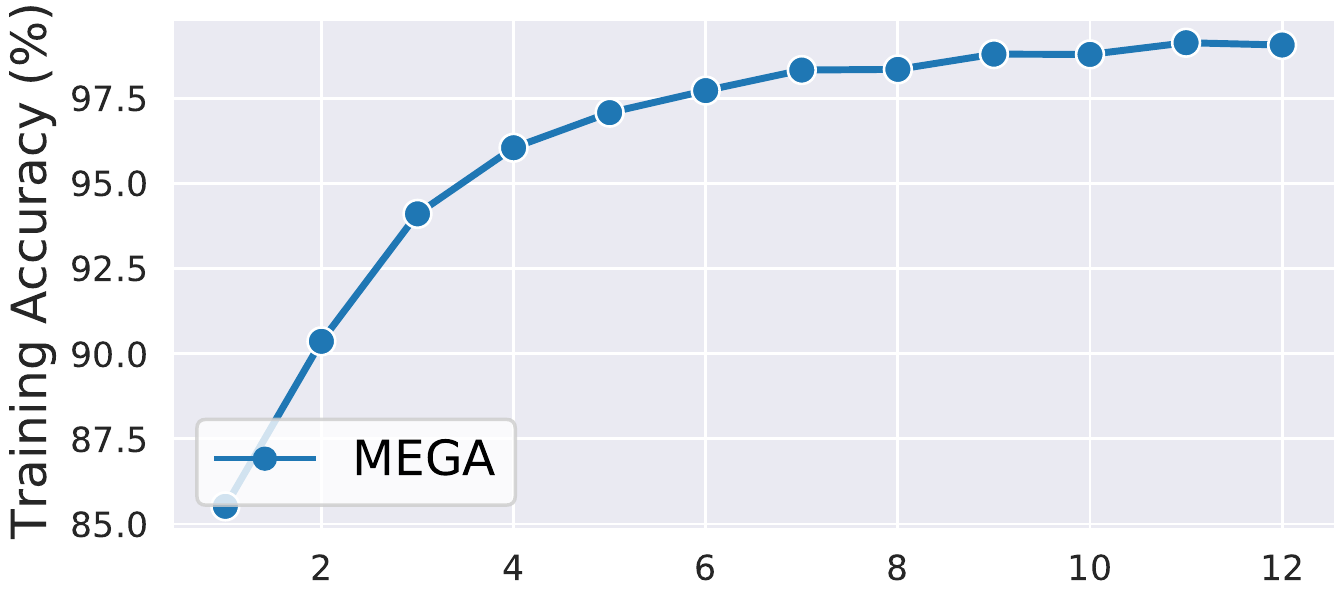} }
\hfill
    \subfloat[Probability-only:MNIST]{
	    \label{subfig:ms_loss_dast}
	    \includegraphics[width=0.35\textwidth]{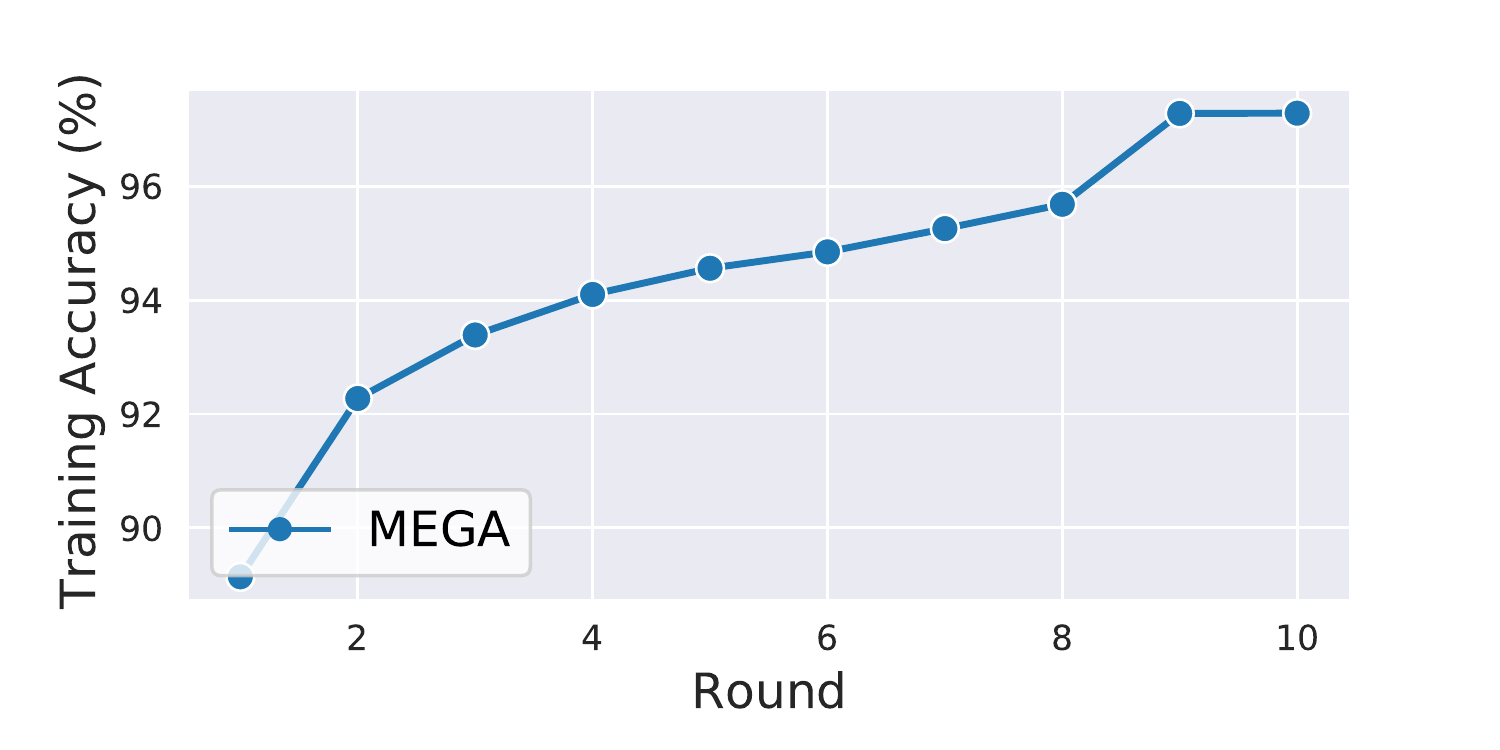}}
	}
	\caption{The training accuracy of the substitute model.}
	\label{supp:trainacc}
\end{figure}

In each round $t$, \alg traverses $(X, \mathcal{T}(X))$ multiple times to train $\mathcal{S}$ until the loss doesn't significantly decrease anymore. Then the local optima of $\mathcal{S}$ for $(X, \mathcal{T}(X))$ can be found and $\mathcal{S}$ can mimic the prediction of $\mathcal{T}$ on the current synthetic examples $X$ with high accuracy. The prediction similarity between $\mathcal{S}$ and $\mathcal{T}$ can be evaluated by the training accuracy of $\mathcal{S}$ on $(X, \mathcal{T}(X))$. We show the evaluation results in Figure~\ref{supp:trainacc}. The training accuracy of $\mathcal{S}$ is high in every round and it continuously increases during the training. Therefore,$\mathcal{S}$ mimics $\mathcal{T}$ well and the assumption is reasonable.

\section{Cross entropy and KL-divergence}

In this section, we demonstrate that cross entropy is equivalent to KL-divergence in the black-box setting. And we also show that cross entropy can avoid gradient vanishing in white box cases where the target model $\mathcal{T}$ is derivable.

Let $\mathcal{S}(x)$ and $\mathcal{T}(x)$ to be the softmax outputs of the substitute model and the target model. The expressions of the KL-divergence loss and the cross entropy loss are shown in the following.

\begin{equation*}
\mathcal{L}_{\mathrm{KL}}(x)=\sum_{i=1}^{N} \mathcal{T}_{i}(x) \log \left(\frac{\mathcal{T}_{i}(x)}{\mathcal{S}_{i}(x)}\right)
\end{equation*}
\begin{equation*}
\mathcal{L}_{\mathrm{CE}}(x)=-\sum_{i=1}^{N} \mathcal{T}_{i}(x) \log \mathcal{S}_{i}(x)
\end{equation*}

In the black-box setting of this paper, $\mathcal{T}$'s model parameters are unavailable. Therefore, in \alg, $\mathcal{T}(x)$ is regarded as a constant vector which is non-differentiable. In this case, we have 

\begin{equation*}
\begin{aligned}
    \nabla_{x} \mathcal{L}_{\mathrm{KL}}(x) = \nabla_{x} \mathcal{L}_{\mathrm{CE}}(x) = - \sum_{i=1}^{N} \mathcal{T}_{i}(x) \nabla_{x} \left[\log \mathcal{S}_{i}(x)\right]
\end{aligned}
\end{equation*}

But an interesting observation is that in white-box settings ($\mathcal{T}$ is derivable) cross entropy won't suffer from vanishing gradients but KL-divergence can suffer from vanishing gradients. The justification is in the following. Taking the gradient over $\mathcal{L}_{\mathrm{KL}}(x)$, we have

\begin{equation*}
\begin{aligned}
\nabla_{x} \mathcal{L}_{\mathrm{KL}}(x) 
&= \sum_{i=1}^{N} \frac{\partial \mathcal{T}_{i}}{\partial x}\log \frac{\mathcal{T}_{i}}{\mathcal{S}_{i}} + \frac{\partial \mathcal{T}_{i}}{\partial x} -\frac{\partial \mathcal{S}_{i}}{\partial x} \frac{\mathcal{T}_{i}}{\mathcal{S}_{i}}\\
&=\sum_{i=1}^{N} \frac{\partial \mathcal{T}_{i}}{\partial x} \log \frac{\mathcal{T}_{i}}{\mathcal{S}_{i}}-\frac{\partial \mathcal{S}_{i}}{\partial x} \frac{\mathcal{T}_{i}}{\mathcal{S}_{i}},
\end{aligned}
\end{equation*}
where $\sum_{i=1}^{N} \frac{\partial \mathcal{T}_{i}}{\partial x} = 0$ because $\sum_{i=1}^{N} \mathcal{T}_{i}=1$. When $\mathcal{S}$ converges to $\mathcal{T}$ we have $\mathcal{T}_{i}(x)=\mathcal{S}_{i}(x)\left(1+\epsilon_{i}(x)\right)$ where $\epsilon_{i}(x)$ tends to $0$ during the convergence process. When $\epsilon_{i}(x)$ tends to $0$, $\log \left(1+\epsilon_{i}(x)\right) \approx \epsilon_{i}(x)$. Then we have

\begin{equation}
\label{kl_van_grad}
\begin{aligned}
\nabla_{x} \mathcal{L}_{\mathrm{KL}}(x)
& \approx \sum_{i=1}^{N} \frac{\partial \mathcal{T}_{i}}{\partial x} \epsilon_{i}-\frac{\partial \mathcal{S}_{i}}{\partial x}\left(1+\epsilon_{i}\right) \\
& \approx \sum_{i=1}^{N} \epsilon_{i}\left(\frac{\partial \mathcal{T}_{i}}{\partial x}-\frac{\partial \mathcal{S}_{i}}{\partial x}\right),
\end{aligned}
\end{equation}
where we have applied $\sum_{i=1}^{N} \frac{\partial \mathcal{S}_{i}}{\partial x} = 0$ because $\sum_{i=1}^{N} \mathcal{S}_{i}=1$. According to Equation~(\ref{kl_van_grad}), the gradient of $\mathcal{L}_{\mathrm{KL}}(x)$ will gradually vanish after many iterations. Then taking the gradient over $\mathcal{L}_{\mathrm{CE}}(x)$, we have
\begin{equation*}
\begin{aligned}
\nabla_{x} \mathcal{L}_{\mathrm{CE}}(x) 
&= -\sum_{i=1}^{N} \frac{\partial \mathcal{T}_{i}}{\partial x}\log \mathcal{S}_{i} + \frac{\partial \mathcal{S}_{i}}{\partial x} \frac{\mathcal{T}_{i}}{\mathcal{S}_{i}}\\
& \approx -\sum_{i=1}^{N} \frac{\partial \mathcal{T}_{i}}{\partial x}\log \mathcal{S}_{i} -\sum_{i=1}^{N} \frac{\partial \mathcal{S}_{i}}{\partial x} - \sum_{i=1}^{N}\epsilon_{i}\frac{\partial \mathcal{S}_{i}}{\partial x} \\
&= -\sum_{i=1}^{N} \frac{\partial \mathcal{T}_{i}}{\partial x}\log \mathcal{S}_{i}  - \sum_{i=1}^{N}\epsilon_{i}\frac{\partial \mathcal{S}_{i}}{\partial x}
\end{aligned}
\end{equation*}
where the first term won't vanish during the iterations. Therefore the cross entropy loss won't suffer from vanishing gradients. Note that we have applied $\sum_{i=1}^{N} \frac{\partial \mathcal{S}_{i}}{\partial x} = 0$.

\section{Applying the baseline DFME}

\subsection{The white-box optimization goal}
We use DFME~\cite{truong2021data} as a baseline in our black-box setting. Here, we briefly introduce the method using the notations of this paper. 

\begin{figure}
	\centering
	    \includegraphics[width=0.48\textwidth]{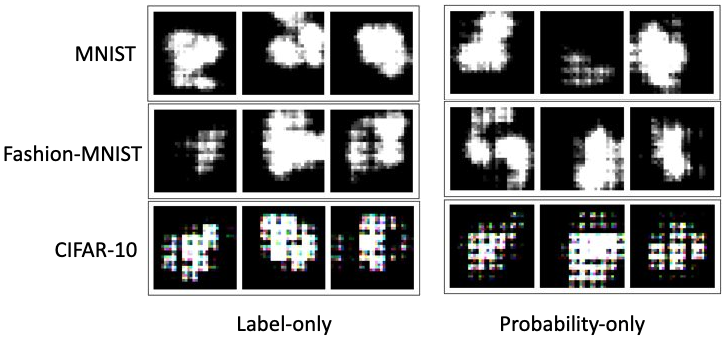}
	\caption{Visualization of generated examples by \alg.} 
	  \label{img:gen_visual}
\end{figure}

DFME makes use of the $\ell_{1}$ norm between the outputs of the target model and the substitute model as the loss function.
\begin{equation*}
   \mathcal{L}(x)=\sum_{i=1}^{N}\left|\mathcal{T}_{i}(x)-\mathcal{S}_{i}(x)\right|,
\end{equation*}
where $x = \mathcal{G}(z)$ is a synthetic example and $N$ is the number of classes.

There is also one generator $\mathcal{G}$ in the method for generating synthetic examples to query the target model. The method follows the principle of generative adversarial networks (GANs). The optimization objectives of $\mathcal{G}$ and $\mathcal{S}$ are competitive. The objectives are show in the following:
\begin{equation*}
    \min _{\mathcal{S}} \max _{\mathcal{G}} \mathbb{E}_{z \sim \mathcal{N}(0,1)}[\mathcal{L}(\mathcal{T}(\mathcal{G}(z)), \mathcal{S}(\mathcal{G}(z)))]
\end{equation*}

The optimization objectives can be applied in white-box settings where the model parameters of $\mathcal{T}$ are accessible.

\subsection{Gradient approximation}
\textbf{The challenge of applying the method in black box settings:} 
In our black-box setting, the target model $\mathcal{T}$ only provides black-box access which means the model parameters of $\mathcal{T}$ are unavailable. If we want to train the generator $\mathcal{G}$, $\nabla_{\theta_{g}} \mathcal{L}$ is required. But $\mathcal{T}$ is not derivable in our black-box setting. Therefore we can't calculate the term $\nabla_{\theta_{g}} \mathcal{T}_i(\mathcal{G}(z))$ where $i = 1,...,N$ and the term is necessary to calculate $\nabla_{\theta_{g}} \mathcal{L}$.

\begin{figure}[t]
	\centering
	    \includegraphics[width=0.48\textwidth]{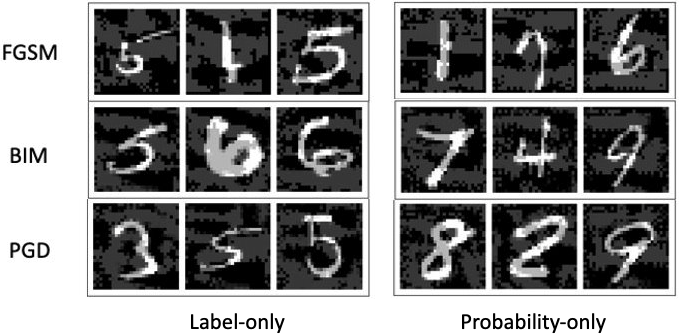}
	\caption{Visualization of adversarial examples on MNIST.} 
	    \label{img:mnsit_visual}
\end{figure}

\begin{figure}[h]
	\centering
	    \includegraphics[width=0.48\textwidth]{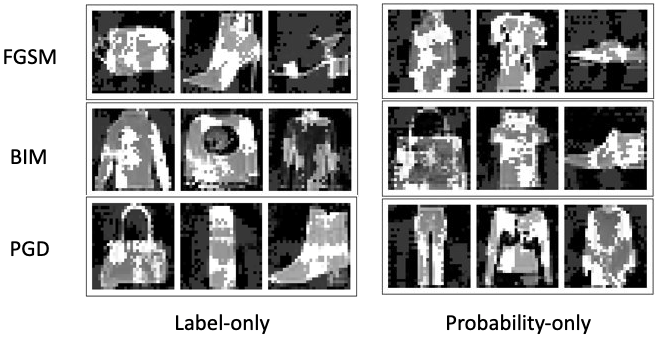}
	\caption{Visualization of adversarial examples on Fashion-MNIST.}
	   \label{img:fashion_visual}
\end{figure}

\begin{figure}[h]
	\centering
	    \includegraphics[width=0.48\textwidth]{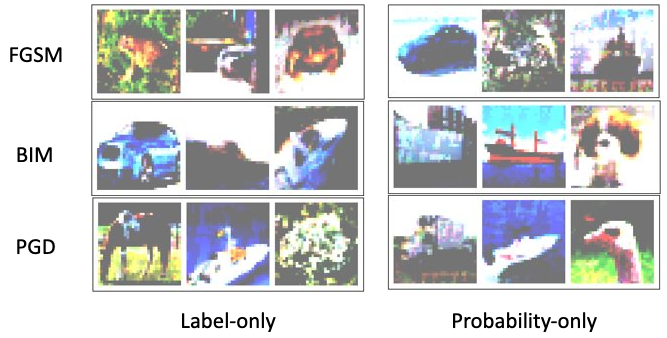}
	\caption{Visualization of adversarial examples on Cifar-10.} 
	  \label{img:cifar_visual}
\end{figure}

\textbf{The solution:} In order to address the aforementioned problem, DFME finds a method to approximate $\nabla_{\theta_{g}} \mathcal{L}$. We know that 
\begin{equation*}
    \nabla_{\theta_{g}} \mathcal{L}=\frac{\partial \mathcal{L}}{\partial \theta_{g}}=\frac{\partial \mathcal{L}}{\partial x} \times \frac{\partial x}{\partial \theta_{g}},
\end{equation*}
where the second term can be computed because $\mathcal{G}$ is derivable and we know it's parameters. Thus, we only need to approximate $\frac{\partial \mathcal{L}}{\partial x}$. This can be done by the forward differences method (Polyak, 1987). The method approximates the gradient of a function at a point by computing directional
derivatives along some random directions which can be formulated as the following equation. For a synthetic example $x \in \mathbb{R}^{d}$, we have the approximate gradient:
\begin{equation*}
    \hat{\nabla}_{x} \mathcal{L}\left(x\right)= \frac{1}{M}\sum_{i=1}^{M}\frac{d \cdot\left(\mathcal{L}\left(x+\epsilon u_{i}\right)-\mathcal{L}(x)\right)}{\epsilon} u_{i}
\end{equation*}
where $u_{i}$ is a random direction (a $d$ dimensional unit vector) and $M$ is the number of directions used for the approximation. The approximate value will be more precise as $M$ increases. Using the solution, DFME can do model stealing in our black-box setting.

\begin{figure*}[t]
	\centering
	{
	\subfloat[Probability-only:MNIST]{
	    \includegraphics[width=0.95\textwidth]{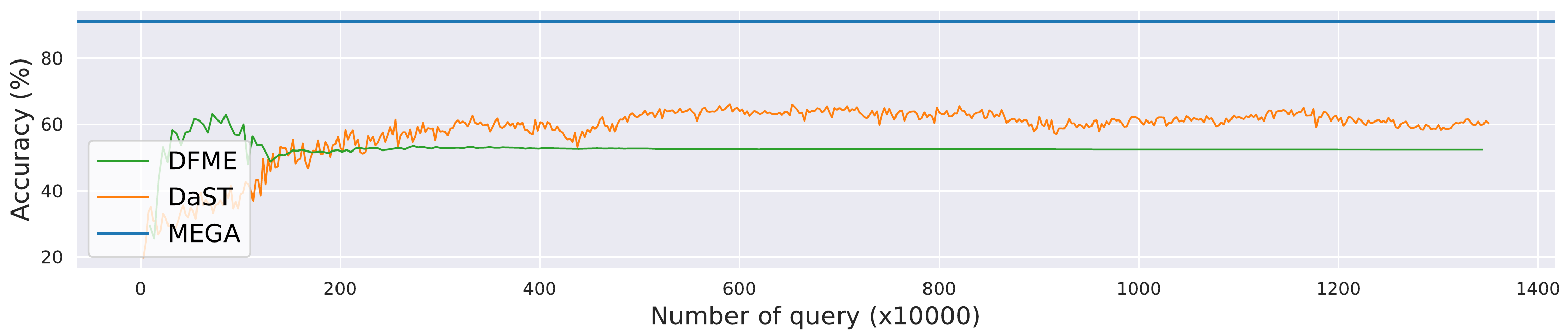} }
\hfill
    \subfloat[Label-only:MNIST]{
	    \includegraphics[width=0.95\textwidth]{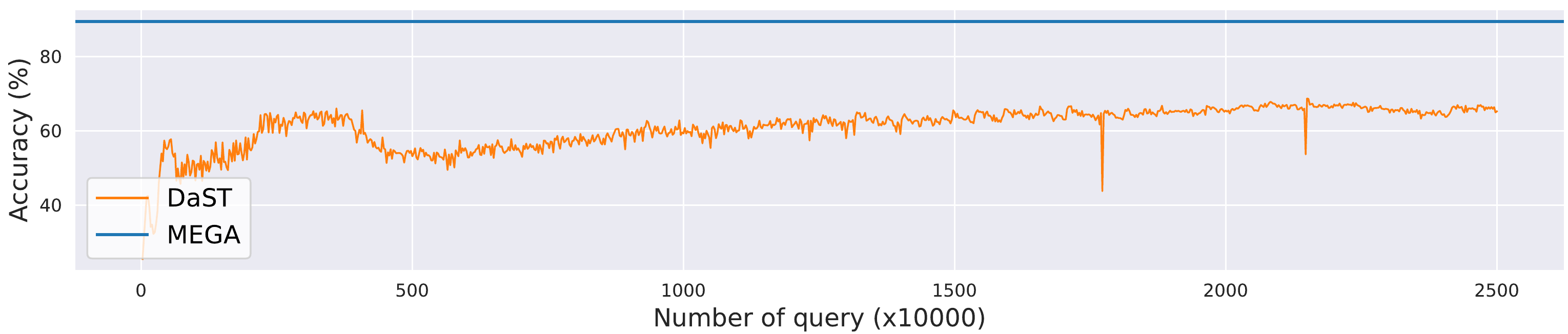}}
\hfill
    \subfloat[Probability-only:Fashion-MNIST]{
	    \includegraphics[width=0.95\textwidth]{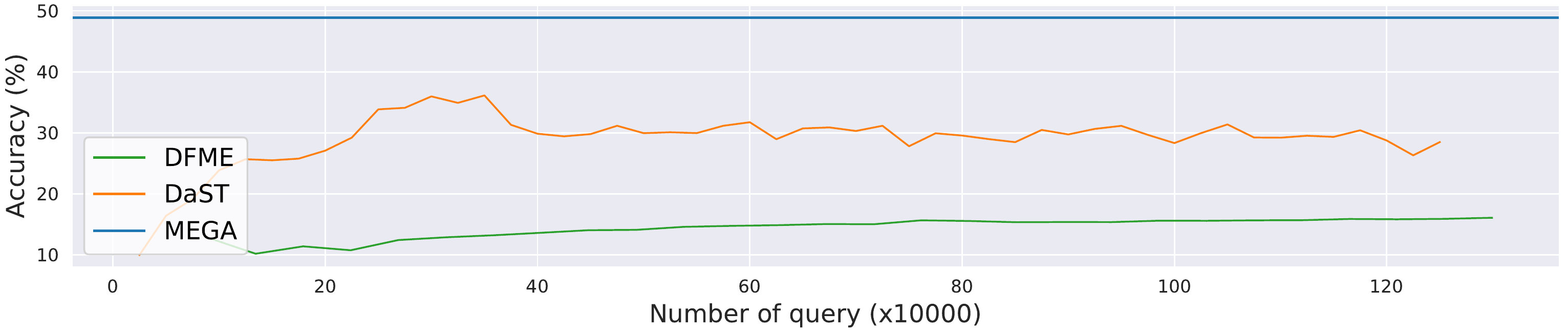}}
\hfill
    \subfloat[Label-only:Fashion-MNIST]{
	    \includegraphics[width=0.95\textwidth]{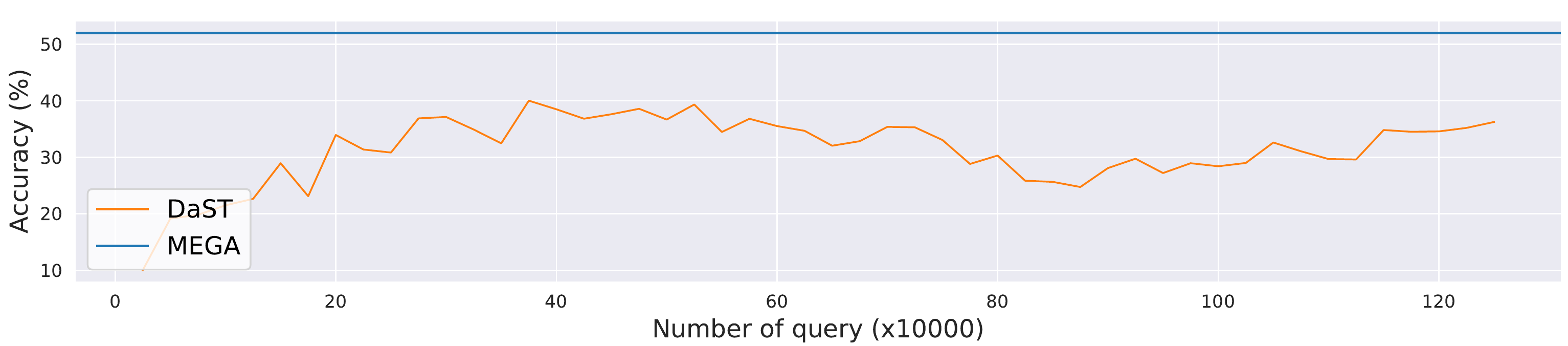}}
	}
	\caption{Substitute model accuracy.}
	\label{fig:full_process}
	\vspace{-1em}
\end{figure*}

\textbf{Weakness:} There are two points of weakness for the baseline. (i) This baseline can only be applied in the probability-only scenario because we need the output probability vector of $\mathcal{T}$ to approximate $\nabla_{\theta_{g}} \mathcal{L}$. (ii) For each synthetic example $x$, this baseline needs additional $M$ queries to approximate the corresponding gradient~\footnote{According to the DFME paper, $M=1$ in the experiments.}.

\section{Additional experimental results}

In this section, we show the model stealing accuracy of DFME and DaST during their training process in Figure~\ref{fig:full_process}. The accuracy of DFME and DaST fluctuates and stays below our approach.

\section{Visualization}

In this part, we visualize the outputs of the generator of our proposed \alg for MNIST, Fashion-MNIST and Cifar-10 datasets in Figure.~\ref{img:gen_visual}. It can be visualized as 3-channel chromatic images for Cifar-10 with 1-channel grayscale images for the others. Adversarial examples generated by our trained substitute model in both label-only and probability-only scenarios are also provided in Figure.~\ref{img:mnsit_visual},~\ref{img:fashion_visual} and ~\ref{img:cifar_visual}.

In Figure.~\ref{img:mnsit_visual},~\ref{img:fashion_visual} and ~\ref{img:cifar_visual}, small perturbations have been added to the original data examples by the state-of-the-art adversarial attack models FGSM, BIM and PGD. We apply small perturbations to generate the adversarial examples.